\documentclass[12pt,preprint]{aastex}

\usepackage{amssymb}
\usepackage{amsmath}
\usepackage{xspace}

\begin{document}

\title{A Unified Monte Carlo Treatment of Gas-Grain Chemistry for Large Reaction Networks. II. A Multiphase Gas-Surface-Layered Bulk Model}

\author{A.I. Vasyunin}
\affil{Department of Chemistry, The University of Virginia, Charlottesville, Virginia, USA}\email{anton.vasyunin@gmail.com}
\author{Eric Herbst}
\affil{Departments of Chemistry, Astronomy, and Physics, The University of Virginia, Charlottesville, Virginia, USA}\email{eh2ef@virginia.edu}

\begin{abstract}
The observed gas-phase molecular inventory of hot cores is believed to be significantly impacted by the products of chemistry in interstellar ices. In this study, we report the construction of a full macroscopic Monte Carlo model of both the gas-phase chemistry and the chemistry occurring in the icy mantles of interstellar grains.  Our model treats icy grain mantles in a layer-by-layer manner, which incorporates laboratory data on ice desorption correctly. The ice treatment includes a distinction between a reactive ice surface and an inert bulk.  The treatment also distinguishes between zeroth and first order desorption, and includes the entrapment of volatile species in more refractory ice mantles.
We apply the model to the investigation of the chemistry in hot cores, in which a thick ice mantle built up during the previous cold phase of protostellar evolution undergoes surface reactions and is eventually evaporated.  For the first time, the impact of a detailed multilayer approach to grain mantle formation on the warm-up chemistry is explored. The use of a multilayer ice structure has a mixed impact on the abundances of organic species formed during the warm-up phase. For example, the abundance of gaseous HCOOCH$_{3}$ is lower in the multilayer model than in previous grain models that do not distinguish between layers (so-called ``two phase'' models). Other gaseous organic species formed in the warm-up phase are affected slightly. Finally, we find that the entrapment of volatile species in water ice can explain the two-jump behavior of H$_{2}$CO previously found in observations of protostars.
\end{abstract}


\section{Introduction}
Interstellar molecules play an important role in the  interstellar medium (ISM) and exist both in the interstellar gas and in ice mantles on interstellar grains \citep[e.g.,][]{HerbstvanDishoeck09}. While many observed molecules are formed via gas-phase chemical reactions \citep{HerbstKlemperer73}, there are important exceptions  such as molecular hydrogen \citep{GouldSalpeter63} and complex organic molecules widely observed in warm interstellar environments \citep[e.g.,][]{GarrodHerbst06}.   Even abundances of simple gas-phase species such as N$_{2}$H$^{+}$, HCO$^{+}$ and CO in some cases may be controlled by chemistry in interstellar ices \citep{Vasyunina_ea12}.

Interstellar ices have  been investigated extensively during the last four decades since the discovery of the 3.1~$\mu$m solid water infrared absorption band in Orion~BN/KL by \cite{GillettForrest73}.   Later observations with ground-based, airborne, and space telescopes (ISO, Spitzer) revealed other main constituents of interstellar ices to be CO \citep{Lacy_ea84}, CH$_{3}$OH \citep{Grim_ea91}, CH$_{4}$ \citep{Lacy_ea91}, CO$_{2}$ \citep{deGraauw_ea96}, NH$_{3}$, and H$_{2}$CO \citep{vanDishoeck04}. The most abundant ice compound in cold sources is H$_{2}$O, with a fractional abundance with respect to H$_{2}$ of $\sim$10$^{-4}$, which is comparable to the concentration of CO, the most abundant gas-phase molecule. The next most abundant ice compounds are CO and CO$_{2}$, with abundances relative to solid water of 0.1---0.3.  Other species such as NH$_{3}$, CH$_{4}$, and H$_{2}$CO have abundances up to a few percent with respect to water \citep{Gibb_ea04}. A comparison of ice inventories around high-mass and low-mass protostars shows a  qualitative similarity of the ice composition in both types of objects.   Quantitatively, however, the  ice composition is somewhat different.  In particular, the fraction of CO and CO$_{2}$ in the ices around high-mass protostars is lower than in ices around low-mass objects \citep{Oeberg_ea11}, possibly reflecting the volatility of these two species.

Analysis of the shapes of absorption bands also yields constraints on the structure of interstellar ices and their evolution during the development of a protostar. Thick ice mantles typically consist of two major components: a water-rich polar ice and a water-poor apolar ice  \citep{Tielens_ea91}.   While CO$_{2}$ normally resides in the polar component, the  CO ice lies in the polar water-poor component.  At late stages of star formation, the CO$_{2}$:H$_{2}$O mixture is  segregated forming CO$_{2}$ ``islands'' within the water ice, while the apolar CO component is gradually desorbed into the gas \citep{Pontoppidan_ea08, Oeberg_ea11}.

Ices manifest themselves in observations of gas-phase molecules in various star-forming environments.   Observations of Class~0 and Class~I protostars performed by \cite{Jorgensen_ea02} and \cite{Schoeier_ea02} indicate that the gas-phase abundance of the simple diatomic molecule CO might require entrapment of CO molecules in water ice or other relatively refractory components of the ice \citep[][]{Cuppen_ea11}. Work by \cite{Ceccarelli_ea01} and \cite{Maret_ea04} show that formaldehyde H$_{2}$CO might be trapped in the ice as well. In hot cores, which represent a later stage of protostellar evolution, complex organic molecules are commonly observed \citep[e.g.,][]{Bottinelli_ea04, Bottinelli_ea07, Bisschop_ea07}.
The organic chemistry in hot cores is driven by heating and complete evaporation of ices formed during the previous cold stage of prestellar evolution \citep[e.g.,][]{Brown_ea88,Caselli_ea93,GarrodHerbst06}. Some organic molecules, such as methanol,  are formed primarily on  ices and then evaporate  into the gas.  For other molecules such as formic acid or methyl formate,  surface formation routes are assisted by gas-phase chemistry, which is activated by chemical precursors ejected from ices \citep[e.g.,][]{GarrodHerbst06}.

The observational data on ices has given impetus to laboratory studies on interstellar ice analogs. Many  studies done during the last decade were directed towards an understanding of the dynamics of ice mixtures in TPD (temperature programmed desorption) experiments \citep{Collings_ea04, Oeberg_ea07, Oeberg_ea09a} and the interaction of ices with UV photons.  Experiments by \cite{Collings_ea04} revealed different patterns of desorption for light volatile species (N$_{2}$, O$_{2}$, CO, CH$_{4}$ etc.) and heavy species with high desorption energies (H$_{2}$O, NH$_{3}$, CH$_{3}$OH). In general, the desorption behavior of species in a mixed ice is complex, and depends on ice composition and thickness, as well as on the heating rate and any phase change in the ice during heating. With a high degree of simplification, one can say that light species can be trapped within more refractory ice, and exhibit at least a two-stage desorption consisting of desorption from the ice surface at their sublimation temperature, and co-desorption of entrapped volatile species with more refractory ice. Refractory species exhibit a much simpler desorption pattern,  corresponding to their temperature of evaporation. In general, the  desorption behavior of species from a thick ice typically shows zeroth-order behavior, in which the rate of desorption doesn't depend on total amount of ice molecules, while for thin ices,  desorption has a first-order nature \citep{SomorjaiBent95}.

The majority of theoretical astrochemical models utilize the treatment of grain surface chemistry proposed in \cite{Hasegawa_ea92} and earlier in \cite{PicklesWilliams77}. This is a very simplified approach in which no details of ice structure and physics is included.   The numerical description of grain surface chemistry in such models is based on chemical rate equations,  which were later proven to have limited applicability to grain-surface chemistry, a problem that can be solved by using computer-intensive stochastic approaches \citep{Vasyunin_ea09}.  There are only a limited number of astrochemical studies where more detailed models of grain surface chemistry have been used. In \cite{HasegawaHerbst93},  rate equations based on a three-phase model consisting of gas, granular surface, and inert bulk were introduced, but rarely used in subsequent studies. \cite{GarrodPauly11} presented an updated version of this model, which includes the treatment of stochastic processes (see below) via a semi-empirical modified rate-equation approach proposed by \cite{Garrod08} and validated in \cite{Garrod_ea09}. \citet[][]{Taquet_ea12} presented a similar model, in which the porosity of interstellar ices is also taken into account, and traditional rate equations are utilized. The important feature of the  three-phase models proposed so far is that, to the best of our knowledge, they are capable of  building up an ice mantle in a layer-by-layer manner, but unable to evaporate it in the same manner, because only abundances of the outermost monolayer are considered. Below this layer, such models only remember average bulk abundances, without distinction between monolayers. The only model of diffusion and desorption of species trapped in a refractory ice is presented in a study of \citet[][]{Fayolle_ea11}, but it does not include any chemistry.

The first serious stochastic models were formulated by  \cite{Charnley98, Charnley01}, who introduced Monte Carlo models of grain surface chemistry based on a stochastic simulation algorithm by \cite{Gillespie76}. In these two papers, gas phase and grain surface chemistries were considered separately. The first Monte Carlo model of coupled gas-grain chemistry based on a large astrochemical network was presented by \cite{Vasyunin_ea09}. In this model, grain surface chemistry was also considered without a distinction between the reactive surface and an inert bulk made of multiple monolayers.

A complex treatment of both the structure and molecular environment in all layers of grain mantles is embodied in models based on a microscopic Monte Carlo approach \citep{Chang_ea05, CuppenHerbst07}.  Here, the motions of individual atoms and molecules on grain ices are modeled. This type of model can include detailed physics of a grain mantle:  layers, porosity, bulk diffusion etc. The main limitation of these models is computational, which renders the incorporation of large chemical networks and simulations of chemistry over long periods of time (more than 10$^{5}$ year) difficult.

The aim of this study is to investigate the chemistry in hot cores affected by evaporation of ices formed during the previous cold stages of star formation with a new model, which includes recent discoveries in the ice physics.  For this purpose, a macroscopic Monte Carlo code  reported in \cite{Vasyunin_ea09} has been modified.  In addition to the correct treatment of stochastic effects and a full gas-grain chemistry, it now includes several important properties of ices: layered structure, entrapment of volatile species in refractory water ice, and layer-by-layer desorption.  With this model, we follow the chemistry starting from ice build-up during the cold collapse phase from semi-diffuse to dense cloud followed by the warm-up stage as a hot core develops.

The outline of the paper is as follows. In Section \ref{modeling}, we describe our chemical and physical models, the algorithm of the model, and its numerical tests. In section \ref{results}, modeling results are presented. Section \ref{discussion} is devoted to a comparison of results with observational data and a discussion. A summary is given in section \ref{summary}.  Finally, the mathematical details of the algorithm used are presented in an Appendix.

\section{Modeling}\label{modeling}

\subsection{Modification of the Monte Carlo algorithm: the MONACO code}\label{algorithm}
In this study, we utilize an extended version of the Monte Carlo approach first described in \cite{Vasyunin_ea09}, which is based on the classical Gillespie algorithm \citep{Gillespie76}. In \cite{Vasyunin_ea09}, the Monte Carlo algorithm was applied to a two-phase astrochemical system consisting of the gas phase and an undifferentiated grain mantle. The fact that a real grain mantle may be thick and consist of multiple monolayers, with just the outermost monolayer (or a few outer layers) chemically reactive while the bulk of grain mantle remains inert,  was not  taken into account.  Here, we extend our algorithm in order to include a more realistic representation of a grain mantle in the model.

Since our approach is macroscopic, we cannot incorporate many realistic microscopic details of the structure of an ice mantle such as surface roughness or bulk porosity,  which can be taken into account in microscopic Monte Carlo simulations. Instead, we consider a very simplified model, although one that is more detailed than previous macroscopic approaches. In our macroscopic Monte Carlo model, molecules are just logical entities that obey certain rules. We assume that: a) an ice mantle can consist of multiple monolayers; b) the formation of the (M+1)th monolayer starts only after the Mth monolayer is fully formed; c)  only a certain number of outer monolayers are chemically active; i.e. chemical reactions  and desorption are only possible between molecules belonging to outer monolayers. The exact number of chemically active monolayers is a subject of calibration of the model vs. laboratory data, as discussed in Section \ref{labbeh};  d) the number of molecules on the grain surface can be increased by accretion from the gas, decreased by desorption into the gas,  and either increased or decreased by a surface chemical reaction depending upon the numbers of reactants or products;  e) there is no porosity or internal motions of molecules in the bulk below the chemically active monolayers; f) only molecules belonging to the chemically active monolayers  can desorb. A detailed description of the algorithm is given in the Appendix.

The important advantage of the developed algorithm is that it not only allows us to {\it grow} the mantle in a
layer-by-layer manner, but makes it possible for it to be evaporated in a layer-by-layer manner as well, contrary to previous three-phase models.
This feature is important for studies of chemistry during the warm-up phase; e.g., in hot corinos and cores.  Because of this feature, we call our model ``multiphase'' in order to make a distinction between our model and previous three-phase models.

The described algorithm was implemented in the MONACO code (acronym for ``the MONte cArlo COde''), which is written in Fortran~90. The code is capable of simulating gas-grain chemistry under time-dependent physical conditions.  Although the code is computationally expensive, it is efficient enough to be run on desktop workstations, albeit  for large periods of time.

\subsection{Behavior of the model vs. laboratory data on ices}\label{labbeh}
It is important to benchmark the proposed algorithm against the results of laboratory studies of ices. The treatment of desorption and entrapment of species in the MONACO algorithm is of special interest, since this part of the algorithm is completely new.  For this purpose, we used experimental results obtained by \cite{Fayolle_ea11}. In that study, TPD experiments involving well-mixed binary and tertiary ice mixtures (CO:H$_{2}$O, CO$_{2}$:H$_{2}$O, CO:CO$_{2}$:H$_{2}$O) were performed with the CRYOPAD experimental setup.  Experiments measured the entrapment efficiency for the volatiles CO and CO$_{2}$ within water ice.  The entrapment efficiency  is here the fraction of CO or CO$_{2}$ with respect to its initial content in ice mixture, which remains trapped within the water ice, after the ice mixture is heated above the CO or CO$_{2}$ sublimation temperature, but below the water sublimation temperature.

It was found that CO and CO$_{2}$ each have two desorption peaks, one of which is caused by desorption of these volatile species from surface layers of ice, and the other  by co-desorption of refractory water ice and entrapped CO and CO$_{2}$ molecules.  The first peaks occur via first-order desorption, while the second peaks indicate zeroth-order kinetics, in which desorbed molecules are replaced by species from within the bulk. The efficiency of entrapment is higher for CO$_{2}$ than for CO. In general, entrapment is more efficient in thick ices (several tens of monolayers) than in thin ice films (6--16 monolayers). A higher fraction of water molecules in ice also increases the efficiency of entrapment. \cite{Fayolle_ea11} also deduced that CO$_{2}$ can diffuse out of the ice bulk to the surface and then evaporate  only from several outermost monolayers of bulk ice, and there is a sharp boundary between monolayers available for  diffusion and the rest of the bulk. Such a complex behavior cannot be fully reproduced within the framework of the simple MONACO algorithm,  which totally neglects diffusion in the bulk ice. To understand the differences between our model and the lab data, we simulated TPD experiment No.~19 from \cite{Fayolle_ea11}, in which desorption vs temperature was studied in a 30~ML (monolayer) tertiary ice mixture of H$_{2}$O:CO$_{2}$:CO (20:1:1). In addition, we compared the results obtained with the MONACO code with an extended three-phase model proposed in \cite{Fayolle_ea11} for the case of a thick 100~ML binary ice with a CO$_{2}$:H$_{2}$O ratio of 1:5. There are no experimental data for such a thick ice in that paper, but the extended three-phase model is based on the whole set of experiments described in \cite{Fayolle_ea11}, and we believe that it gives a reasonable estimation of entrapment efficiency for thick ice. For the case of experiment~19, \cite{Fayolle_ea11} deduce an entrapment efficiency of 84\%, for either the CO or CO$_{2}$ species,  while for the case of thick ice their model gives an entrapment efficiency of 95\% for CO$_{2}$. The second case should also be considered as astrochemically relevant despite its lack of CO, because the thickness of the interstellar ices is estimated to be $\sim$100~ML and the relative fraction of CO and CO$_{2}$ to water varies between 1:10 and 1:3 \citep{Gibb_ea04, Oeberg_ea11}.

\begin{figure}\centering
\includegraphics[width=0.45\textwidth]{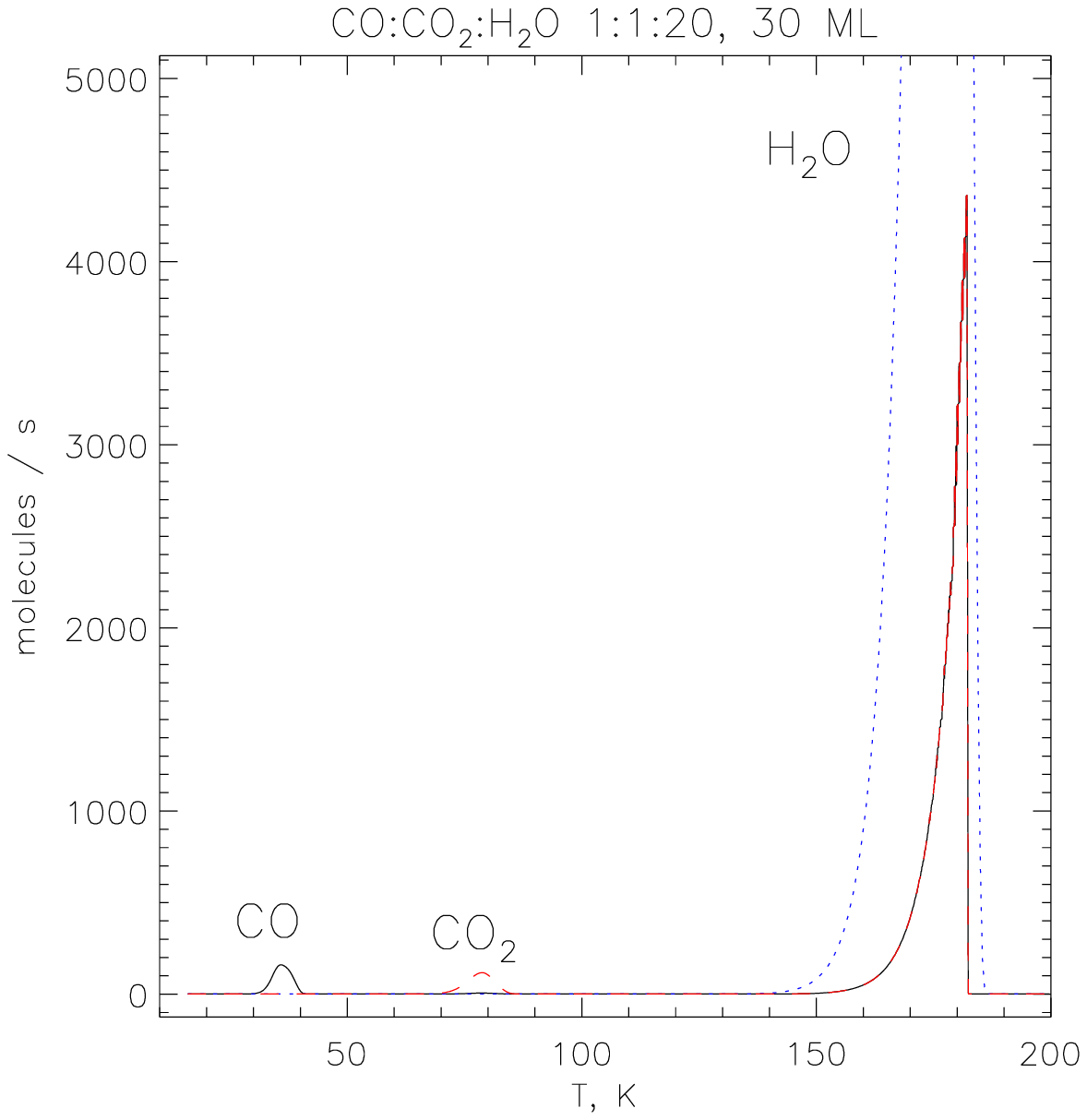}
\includegraphics[width=0.45\textwidth]{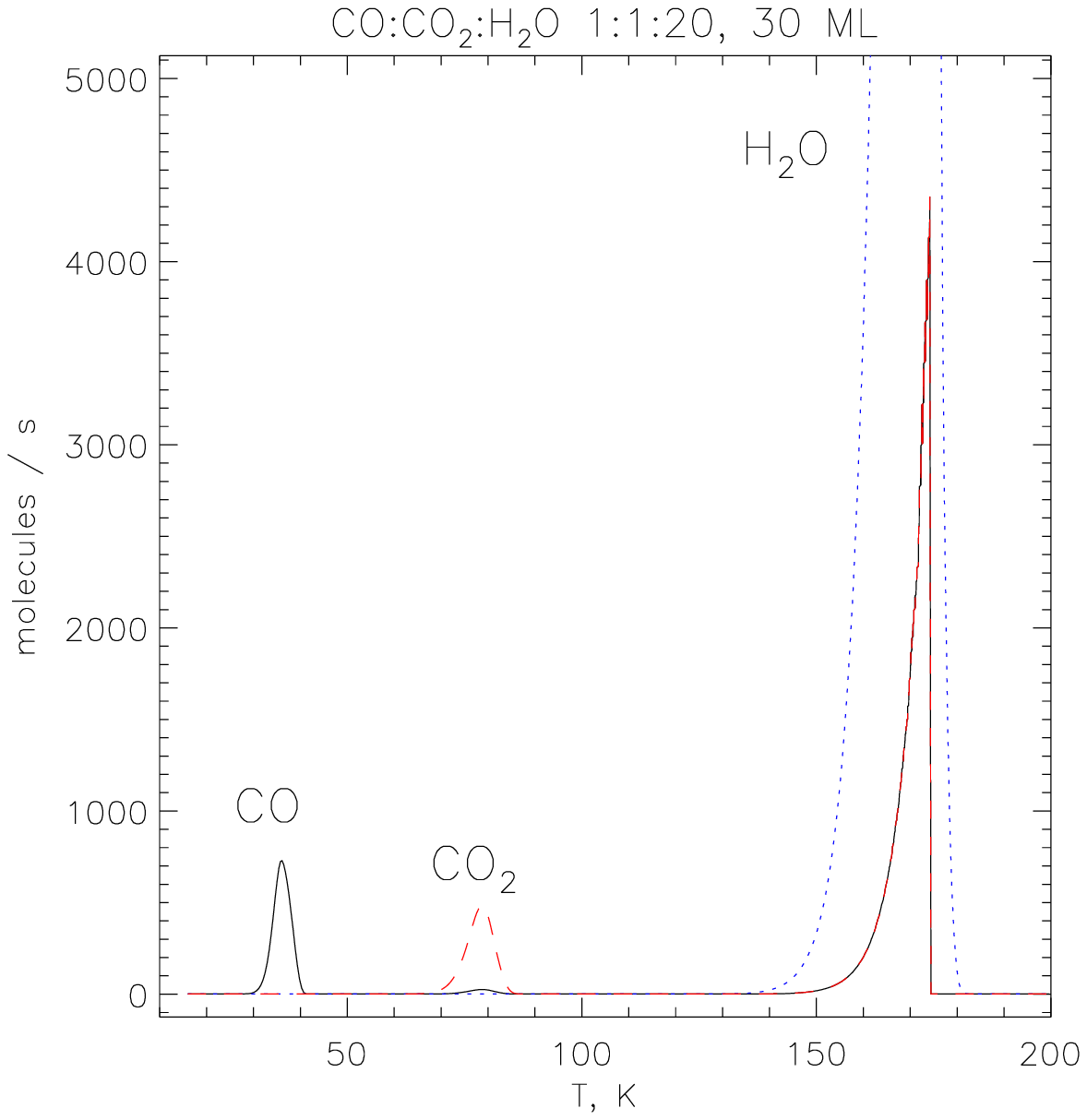}
\caption{MONACO simulation of desorption rate vs. temperature for a TPD experiment with 30 ML tertiary ice of composition H$_{2}$O:CO$_{2}$:CO 20:1:1. Left panel: 1 monolayer is chemically active.  Right panel: 4 monolayers are chemically active. The CO peaks are shown in solid black, the CO$_{2}$ peaks in dashed red, and the water peak in dotted blue.}\label{tpd}
\end{figure}

To model the TPD experiments, we set the initial ice composition in the MONACO model according to those used in corresponding experiments of \cite{Fayolle_ea11}. The temperature in the model is then linearly increased over the timescale used in the experiments. In contrast to actual simulations described later in this study, no chemical reactions are considered.  Therefore, the desorption energies of the species are the only parameters needed.  These energies are 1150~K for CO \citep{GarrodHerbst06}, 2850~K for CO$_{2}$ \citep[][]{GarrodHerbst06} and 5700~K for H$_{2}$O \citep[][]{Fraser_ea01}.
Figure \ref{tpd} represents the results of simulations of the TPD experiment with the  tertiary CO$_{2}$:CO:H$_{2}$O ice mixture performed with the MONACO code in which one outermost monolayer is chemically active (left panel), or four outer monolayers are chemically active (right panel).   From the comparison of the figure with Figure~4a in \cite{Fayolle_ea11}, one can see that the results of the numerical TPD experiment obtained with the MONACO code are qualitatively similar to the real experimental results. Both CO (solid black) and CO$_{2}$(dashed red) exhibit two peaks during the ice warm-up. The first peaks correspond to the evaporation of these volatile species from the ice surface, while the second peaks are caused by the co-desorption of entrapped volatiles and refractory water ice. The shapes of the surface desorption peaks and bulk desorption peaks confirm that the MONACO code treats zeroth- and first-order desorption correctly. The desorption peak of water is located in our model at higher temperatures than in the experiments of  \cite{Fayolle_ea11} because the desorption energy of water we adopt following \cite{Fraser_ea01} is somewhat higher than those found by \cite{Fayolle_ea11} in their experiments.

\begin{figure}\centering
\includegraphics[width=0.45\textwidth]{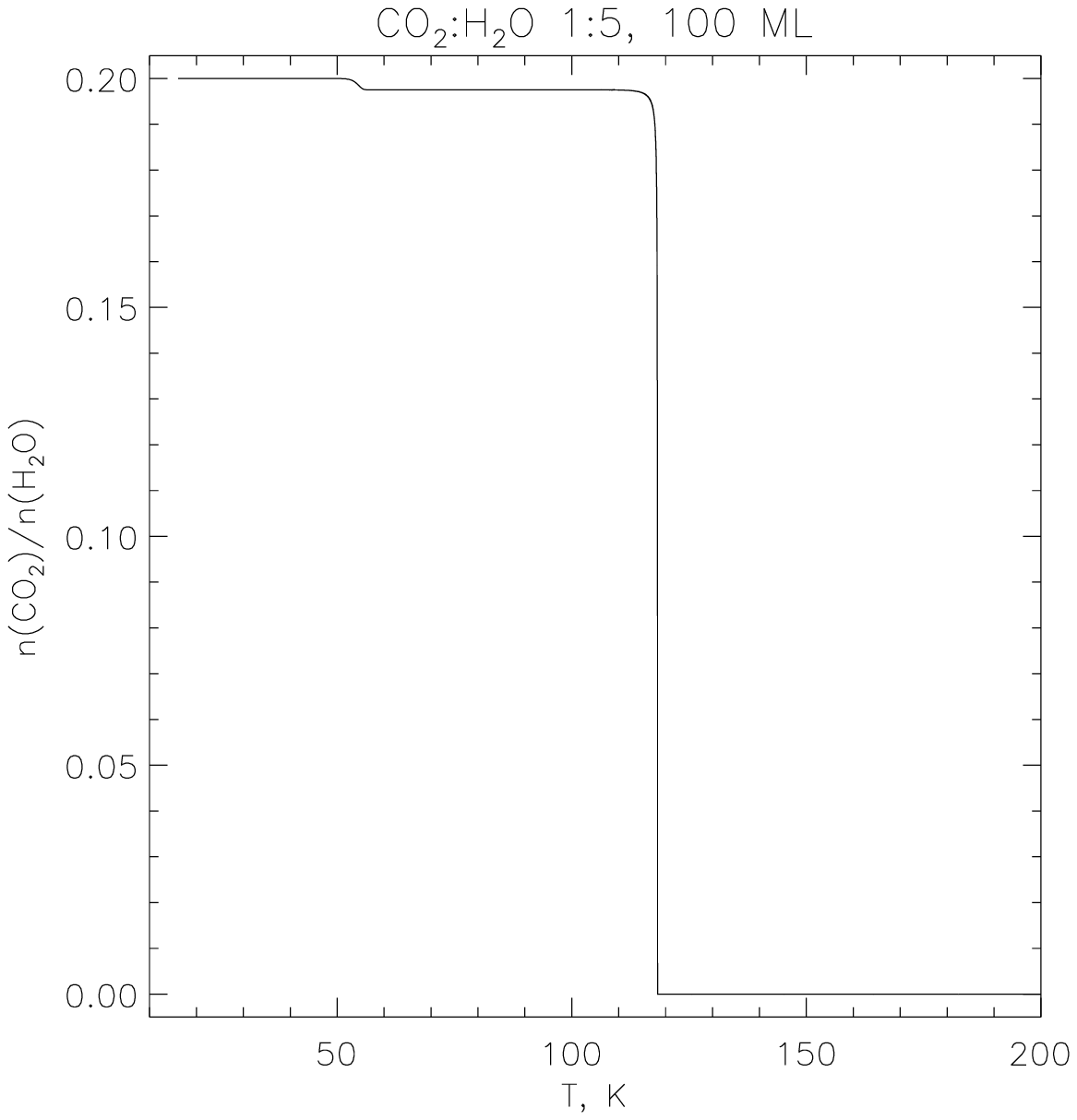}
\includegraphics[width=0.45\textwidth]{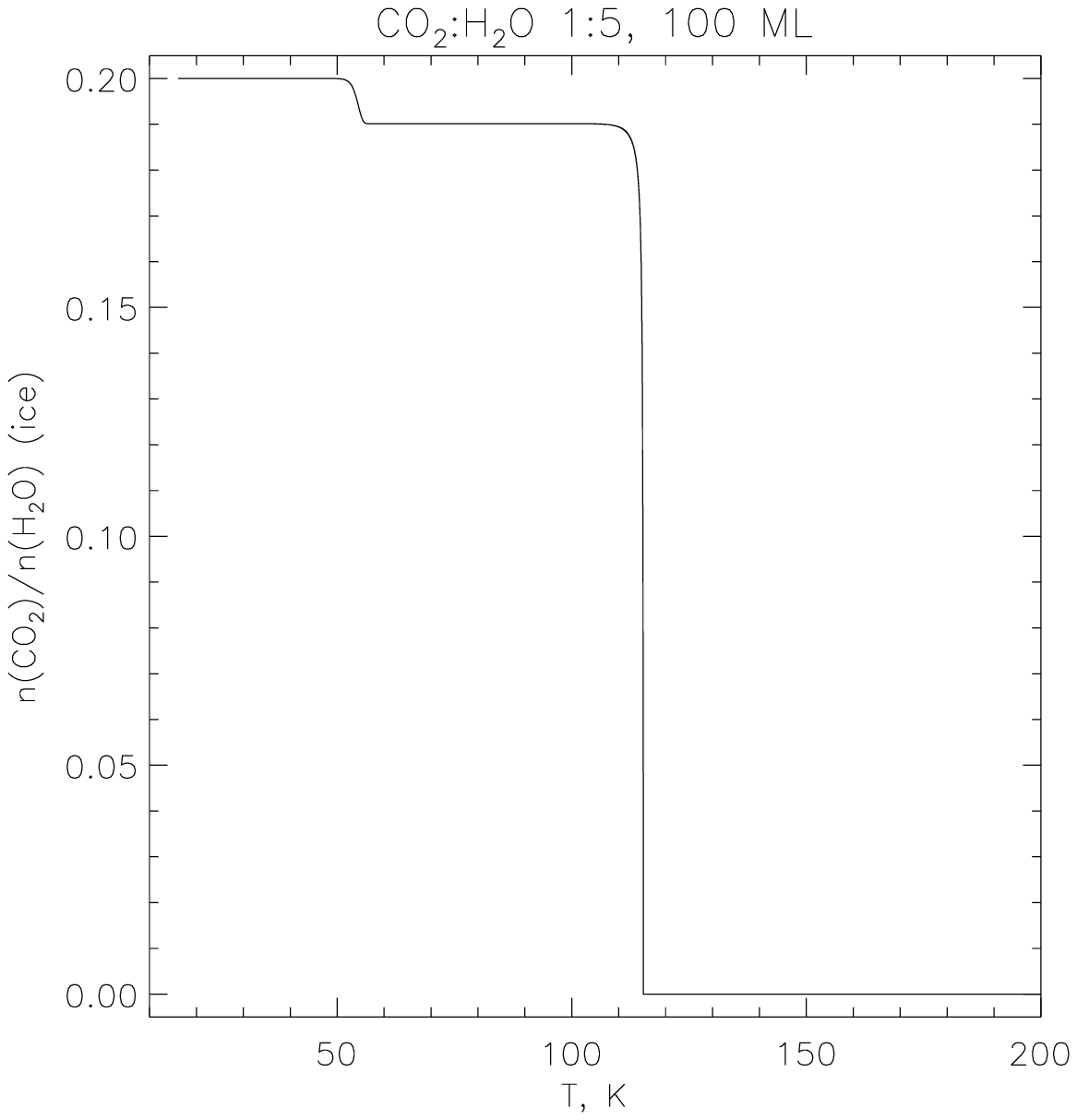}
\caption{CO$_{2}$:H$_{2}$O ratio as a measure of entrapment efficiency in a MONACO simulation of a model by \cite{Fayolle_ea11} with 100 ML of a binary ice mixture of composition CO$_{2}$:H$_{2}$O 1:5. Left panel: 1 monolayer is chemically active.  Right panel:  4 monolayers are chemically active.} \label{enteff}
\end{figure}

Let us now consider the quantitative agreement of laboratory TPD data and TPD simulations with our code. In Figure~\ref{enteff}, the simulated fraction of CO$_{2}$ in the 100~ML CO$_{2}$:H$_{2}$O ice mixture versus temperature is shown vs temperature. Once again, the left panel corresponds to our simulation with one chemically active monolayer, while the right panel represents results obtained  with four chemically active monolayers. Using this plot, it is possible to estimate the entrapment efficiency of CO$_{2}$ in the simulated ice sample. For that, one should divide the value of CO$_{2}$:H$_{2}$O ratio in the temperature range 60--100~K by the value of CO$_{2}$:H$_{2}$O fraction at  temperatures below 50~K. A comparison of the left panel of Figure~\ref{enteff} with Figure~8 in \cite{Fayolle_ea11} reveals  that in our simulation with one outermost chemically active monolayer,  the entrapment efficiency of volatile CO$_{2}$ is significantly higher than in the extended three-phase model of \cite{Fayolle_ea11}. This fact can also be seen when comparing the left upper panel of Figure~\ref{tpd} with Figure~4a in \cite{Fayolle_ea11}, where the relative heights of the surface desorption peaks and co-desorption peaks are smaller in the simulation than in the real TPD experiment. Specifically, the entrapment efficiency in the simulation of the 30~ML tertiary ice CO:CO$_{2}$:H$_{2}$O is 96.3\% vs 84\% in the experiment and 98.8\% vs 95\% for the 100~ML binary CO$_{2}$:H$_{2}$O ice. We conclude that in the absence of bulk diffusion only a very small fraction of volatile species is available for desorption from the ice when the temperature is below the water evaporation threshold.

To overcome this problem and make our modeling results closer to the laboratory data, we changed our initial assumption that only the outermost monolayer of the ice mantle is chemically active. It was found that by making the four upper monolayers of ice mantle chemically active, we can make our modeling results much closer to the results from \cite{Fayolle_ea11} for the astrophysically important case of the 100~ML binary CO$_{2}$:H$_{2}$O ice. This assumption also helps to reproduce the TPD experiment with the 30~ML tertiary ice. The results of our simulations with the tuned-up MONACO code are presented in the right panels of Figs.~\ref{tpd} and \ref{enteff}, and may be compared with Figure 4a and Figure~8 in \cite{Fayolle_ea11}. Our revised values for the entrapment efficiency are 85\% for the tertiary ice and 95.2\% for the binary ice, which lie very close to experimental and model values. It is important to note that our revision of the code  cannot fully compensate for the absence of bulk diffusion. In general, one can expect that the revised code will overestimate the efficiency for the entrapment of volatile species in the case of ices with a low fraction of refractory water (see, e.g., Table~1 in \cite{Fayolle_ea11}). However, for the purpose of this study, in which the evaporation of thick ice with water as the major component is investigated, the adopted revision is reasonable.

\subsection{Chemical model}
In this study, we utilize a full gas-grain chemical model, in which the gas-grain chemical network is a version of the KIDA database published in \cite{Semenov_ea10}\footnote{kida.obs.u-bordeaux1.fr/uploads/models/benchmark\_2010.dat} with several minor updates concerning missing desorption processes for several molecules in the original file. The KIDA network in turn, is a successor of the OSU astrochemical database, which can be found at http://www.physics.ohio-state.edu/$\sim$eric/research.html  \citep{Wakelam_ea12}. In total, the chemical network includes 662 species and 5693 chemical processes. The surface network does not include the many additional complex molecules in the most recent OSU network used by \cite{Garrod_ea08}. Nevertheless, it is sufficient for this work, since in the warm-up phase we limit our study of complex organic species to dimethyl ether (CH$_3$OCH$_3$), formic acid (HCOOH), and methyl formate (HCOOCH$_3$).

There are two major modifications introduced to the original rate file. First, the dissociative recombination channels of protonated dimethyl ether were replaced with those taken from Table~4 of \cite{Hamberg_ea10}. Second, the formation of protonated methyl formate via the gas-phase route CH$_{5}$O$^{+}$ + H$_{2}$CO $\rightarrow$ H$_{5}$C$_{2}$O$_{2}^{+}$ + H$_{2}$ was disabled according to \cite{Horn_ea04}. Also, the desorption energy of molecular hydrogen was chosen to be 314~K according to \cite{Katz_ea99}. The gas-phase and grain surface chemistry are connected by accretion and desorption processes. For accretion, the sticking probability for all neutral species except for H, H$_{2}$ and He is unity. Molecular hydrogen, helium and ions are not allowed to stick on the grain surface, although selected ion reactions with negatively charged grains are included. The non-sticking of molecular hydrogen in our model is dictated by computational problems.  Including the accretion of molecular hydrogen would slow down our Monte Carlo simulations by a factor of $\sim$10000, and make them unfeasible.  The role of reactions involving radicals and H$_{2}$ on surfaces has been investigated by \citet[][]{HasegawaHerbst93} in a model that uses simple rate equations with a common barrier for all such reactions.  In general, hydrogenation on surfaces is handled efficiently by association reactions with atomic hydrogen. In our network, we produce water via the well-known sequence of reactions between atomic oxygen and atomic hydrogen  \citep[e.g.,][]{Dulieu_ea12}.  As  is shown later in this study,  the water on grain surfaces is produced in abundance by this mechanism at the studied temperatures. Desorption processes enabled in our model include thermal evaporation, cosmic ray induced desorption, and photodesorption. Given the uncertainty of laboratory experiments and lack of experimental data for the majority of molecules, we adopted a photodesorption yield of 10$^{-3}$ for all surface species according to \cite{Oeberg_ea07}. Only molecules in chemically active monolayers are subject to desorption.

We adopted a cosmic ray ionization rate for H$_2$ of $\zeta$=1.3$\cdot$10$^{-17}$~s$^{-1}$. The single granular radius in our model is 10$^{-5}$~cm.  The elemental abundances correspond to set EA1 from \cite{WakelamHerbst08}.  All elements initially are in atomic form except for hydrogen, which is initially bound up in H$_{2}$.

We consider only diffusive surface chemistry via the Langmuir-Hinshelwood mechanism. The surface mobility of species is described using the formalism from \cite{Hasegawa_ea92}. According to this formalism, there are two sources of mobility for species on grain surfaces: thermal hopping and quantum tunneling through diffusion barriers for the lightest species --- H and H$_{2}$. The main control parameters for the diffusive chemistry   ---  the diffusion/desorption energy ratio $E_{\rm b}$/$E_{\rm D}$ and the thickness of the potential barrier for diffusion --- are known poorly. In the literature, adopted values for $E_{\rm b}$/$E_{\rm D}$ vary from 0.3 \citep{Hasegawa_ea92} to 0.77 \citep{RuffleHerbst00}. Tunneling for the lightest species is assumed to be efficient in \cite{Hasegawa_ea92}, but later studies do not confirm that \citep{Katz_ea99}, although some recent models also include it \citep[][]{CazauxTielens04, Iqbal_ea12}. Having in mind these uncertainties, we utilized three sets of control parameters for the surface mobility of species in this study. In the first set, we assume a diffusion/desorption energy ratio of 0.3 and efficient tunneling for H and H$_{2}$ on surfaces through thin rectangular potential barriers of width 1\AA. In the second and third sets, tunneling for light species is switched off, with $E_{\rm b}$/$E_{\rm D}$=0.5 in the second set, and 0.77 in the third set. Therefore, the first and the third sets correspond to extreme values considered before in \cite{Hasegawa_ea92} and \cite{RuffleHerbst00}. The second set represents an intermediate case and was utilized previously in, e.g.,   \cite{GarrodHerbst06}. The three sets of control parameters for the mobility of surface species are identically implemented in the MONACO code and in a standard two-phase rate-equation code (RE), which is used for comparison of the new multilayer approach with traditional models.

For all three sets of control parameters, the rates of all surface reactions that have activation barriers were calculated assuming quantum tunneling through rectangular activation barriers of width 1~\AA~according to Eq.~(6) in \citet[][]{Hasegawa_ea92}. The set of activation barriers of surface reactions originates from the earlier work of \citet[][]{GarrodHerbst06}, which in turn is partially based on the  compilation of sets from \citet[][]{Hasegawa_ea92} and \citet[][]{RuffleHerbst00}. The majority of activation barriers  are taken from analogous gas-phase reactions. This fact motivated our decision not to include a so-called reaction-diffusion competition for surface reactions with activation barriers  in the model \citep[][]{HerbstMillar08}.   Competition only pertains to diffusive surface reactions, not to gas-phase ones. Therefore, the values of activation barriers inherited from measurements of gas-phase reactions may not be valid when reaction-diffusion competition is enabled for surface reactions. Instead, barriers obtained directly for surface reactions should be used \citep[e.g.,][]{Cuppen_ea09, ChangHerbst12}. These data are available only for several reactions, and their incorporation into a full-scale network of reactions is problematic.

We do not use the modified rate equations proposed by \cite{Garrod08} in this study despite their relative success in other studies, because they were formulated only for those two-body grain surface reactions, for which reactants appear on a grain surface independently of each other.
If a surface reaction has products that are reactants for other surface reactions, such as photodissociation, the crucial condition of independence is violated. Further study on this point is warranted.

\subsubsection{Multiple layers of ice mantles}

According to \cite{Hasegawa_ea92}, the mobility of species on grain surfaces can be converted to the rate of a surface two-body reaction between species $i$ and $j$ using the expression:
\begin{equation}\label{rsurf}
r_{\rm ij}=k_{ij}(r_{\rm hop}^{i}/N_{\rm s}+r_{\rm hop}^{j}/N_{\rm s})\langle n_{\rm i}\rangle\langle n_{\rm j}\rangle
\end{equation}
where $r_{\rm hop}$ is the hopping rate or tunneling rate for a species over or through the diffusion barrier to a nearest neighbor site, $k_{ij}$ is a factor $\le 1$ related to tunneling under any activation energy of the reaction, if it exists, $\langle n_{\rm i}\rangle$ and $\langle n_{\rm j}\rangle$ are the average abundances of species $i$ and $j$ on a grain surface, and $N_{\rm s}$ is the total number of adsorption sites on the  surface.

This expression is reasonable when the total number of species on a grain surface $N_{\rm tot}$ does not exceed $N_{\rm s}$. However, if the total number of species on a grain exceeds $N_{\rm s}$, some fraction of molecules in the ice lies buried within the mantle and might not be available for diffusion. In the case of a thick mantle,  the majority of species are buried. Hence, if we assume that only molecules of one surface monolayer are reactive, we must correct  expression (\ref{rsurf}) to
\begin{equation}\label{rscorr}
r_{\rm ij}^{new} = r_{\rm ij}\cdot[min(N_{\rm s}/N_{\rm tot},1)]^2.
\end{equation}
For the case of unimolecular surface photoreactions, if we believe that they occur only in the outermost monolayer, expression (\ref{rscorr}) should be modified to
\begin{equation}\label{rscorrph}
r_{\rm ij}^{new} = r_{\rm ij}\cdot min(N_{\rm s}/N_{\rm tot},1)
\end{equation}
These corrections, in principle, should be implemented in the two-phase rate-equation model. However, we did not include it in in our RE model so that our two-phase model could resemble as closely as possible those used in  previous studies.

Since we assume four upper monolayers to be chemically active according to Section~\ref{labbeh}, but also wish to investigate the impact of corrections (\ref{rscorr}) and (\ref{rscorrph}) on the modeling results, we consider two variations of our MONACO model: MC-1 in which corrections (\ref{rscorr}) and (\ref{rscorrph}) are implemented, and MC-4, in which we omit this correction. In Model MC-1, the corrections lead to the result that when the mantle thickness is four monolayers or more, the rates of unimolecular and two-body surface reactions are slower by a factor of 4 and 16, respectively,  than in model MC-4. Thus, we have three models -- RE, MC-1 and MC-4 -- with physically distinct assumptions, which allow us to investigate the differences among assorted  two-phase and  multiphase models, each of which was run with the three sets of $E_{\rm b}/E_{\rm D}$ discussed previously.

\subsection{Physical model}
\begin{figure*}\centering
\includegraphics[width=0.35\textwidth,angle=90]{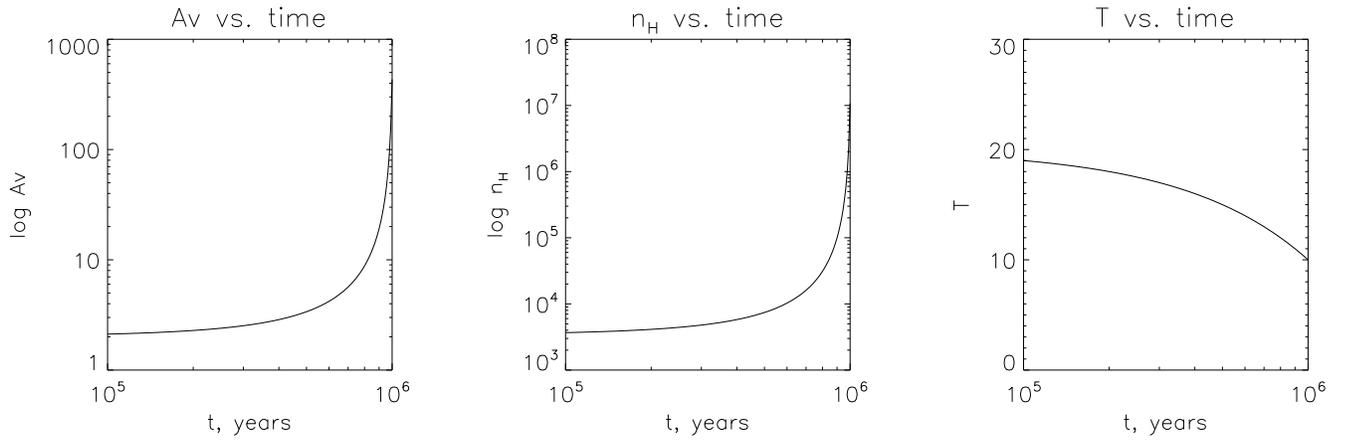}\caption{Key physical parameters of the model vs. time during the cold collapse stage.}\label{modpar}
\end{figure*}

For this study, we adopt a schematic two-stage model of formation of a hot core aiming to mimic the evolution of a gas parcel from the diffuse cloud phase to the hot core phase. The idea of the model is similar to those described in \cite{GarrodHerbst06}, and inherited from earlier
work \citep{Viti_ea04}. The model consists of two stages. In the first ``cold'' stage, a free-fall collapse occurs \citep{Spitzer78,Brown_ea88}. Physically, we attribute the cold collapse phase in our model to the formation of a cold prestellar cloud from a translucent cloud.  The collapse starts at a gas density $n_{\rm H} =3 \times 10^{3}$~cm$^{-3}$ and a visual extinction of A$_{V}$=2 and continues until the desired final density is reached. The collapse continues until a gas density of 10$^{7}$~cm$^{-3}$ is reached after 10$^{6}$ yr of evolution. Figure~\ref{modpar} shows the change in the visual extinction, gas density, and temperature during the cold collapse. Despite the fact that  free-fall collapse is isothermal, in our study we assume that  the temperature linearly drops from 20~K to 10~K with time during the collapse. Physically, this corresponds to the fact that the efficiency of radiative heating of grains drops as the visual extinction increases. Given the number of uncertainties in our knowledge of the collapse and of the properties of interstellar grains, we believe that such a schematic approach is sufficient for our model, and a more detailed treatment of the dust temperature \citep{GarrodPauly11} is not necessary. The gas temperature is assumed always to be equal to the dust temperature in our model.

The second stage of our model is the warm-up phase. Here we assume that as the gas and dust collapse towards the protostar,  a quadratic warm-up from 10~K to 200~K over $2 \times 10^{5}$~yr following \cite{Viti_ea04} and \cite{GarrodHerbst06}, which corresponds to  intermediate mass star formation. In the warm-up phase, the density remains constant.

\section{Results}\label{results}

\subsection{Cold Collapse phase: build-up of the ice mantle}

\subsubsection{Early collapse stage}
\begin{figure*}\centering
\includegraphics[height=0.8\textwidth, angle=90]{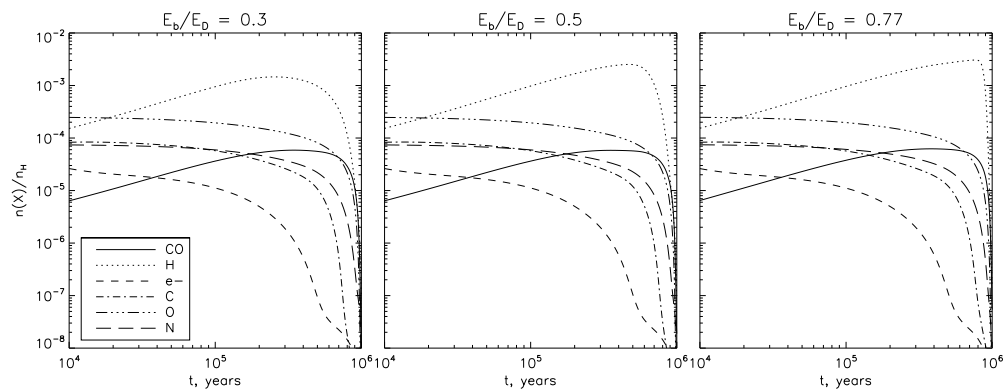}\vspace{5mm}\caption{The evolution of the fractional abundance of simple gas-phase species during the whole cold collapse phase.}\label{all3-1}
\end{figure*}

\begin{figure*}\centering
\includegraphics[height=0.8\textwidth, angle=90]{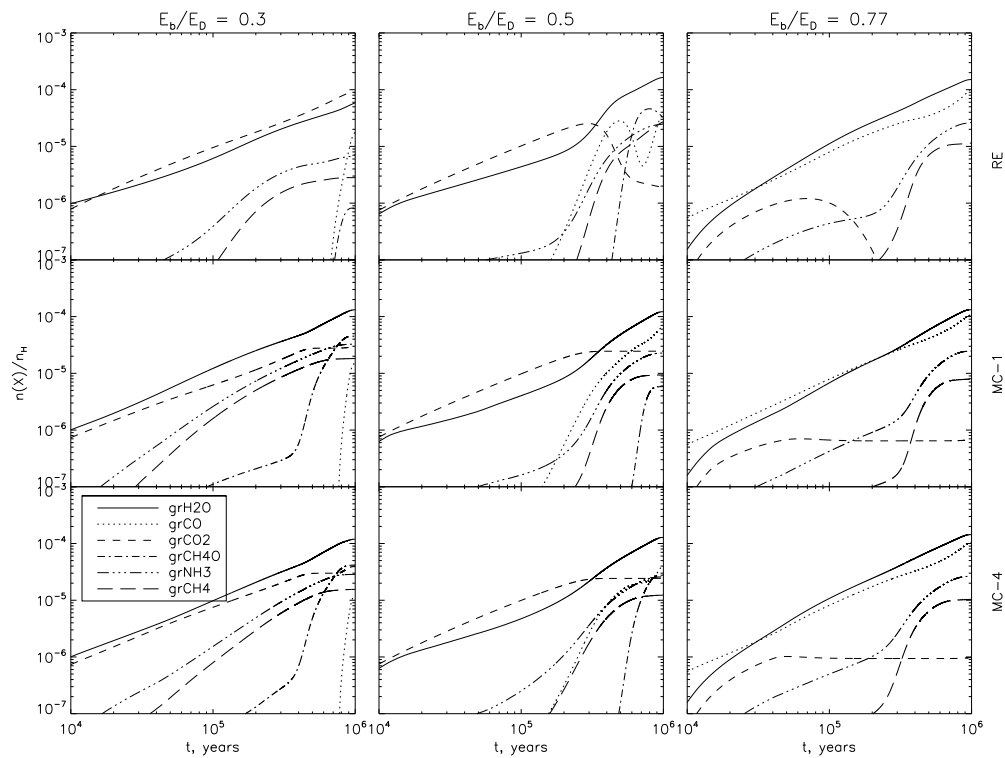}\vspace{5mm}\caption{An overview of the evolution of the ice composition in different models during the whole cold collapse phase. The molecular concentrations are plotted as fractional abundances with respect to the gas-phase nuclear hydrogen density $n_{\rm H}$.}\label{all9}
\end{figure*}

The fractional abundances of selected simple gas-phase species during the whole cold collapse phase are shown in Figure~\ref{all3-1} for Model MC-1. This figure has three panels, one for each $E_{\rm b}$/$E_{\rm D}$ ratio considered.  Note that there is little difference among the abundances in the three panels. The abundances are affected even less if we switch to model RE or MC-4, so nine panels are not required. Figure~\ref{all9} contains fractional abundances for the major ice species as a function of time during the whole cold collapse stage.  Here there are nine panels since differences among the two-phase rate-equation approach (RE) and the multilayer approaches (MC-1 and MC-4) can be seen.

\begin{figure}\centering
\includegraphics[width=0.6\textwidth]{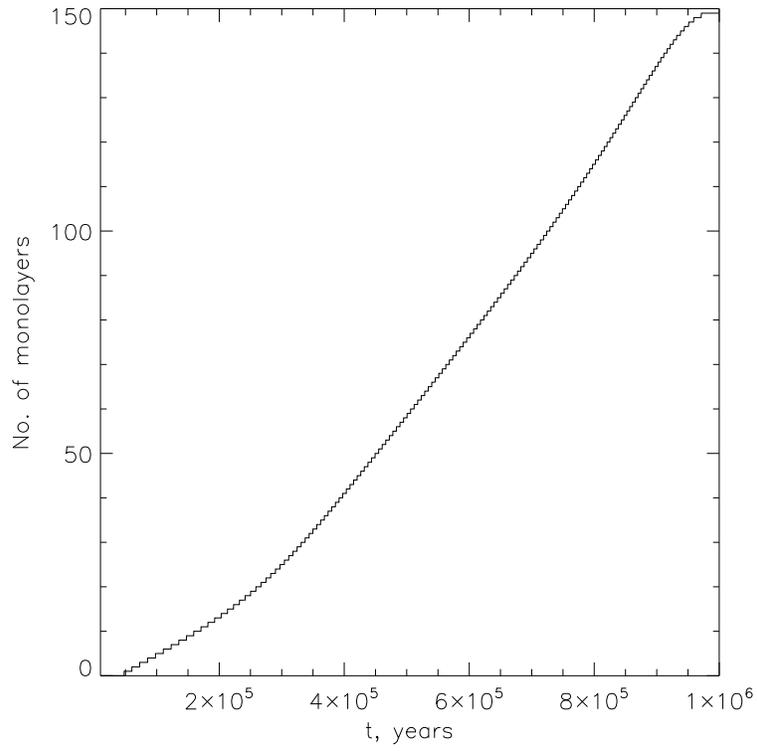}\caption{A typical plot of the calculated ice mantle thickness
vs.~time during the cold collapse phase. The grain surface is bare at the beginning of the simulation.} \label{mantlegrowth}
\end{figure}

During the early stage of collapse  (up to $2 \times 10^{5}$~yr), the chemical evolution of the major ice mantle species is qualitatively similar in the RE, MC-1 and MC-4 models (see each column in  Figure~\ref{all9}). This result is natural, because at  this very early stage the thickness of the ice mantle is $\le$~10 monolayers, as shown in
Figure~\ref{mantlegrowth}, and the number of molecules participating in surface chemistry in similar all models.
On the other hand, the different $E_{\rm b}/E_{\rm D}$ ratios can cause more significant differences in the ice composition, although models with $E_{\rm b}$/$E_{\rm D}$=0.3 and 0.5 are quite similar during this early stage. The two major constituents of the ice in these models are carbon dioxide (CO$_{2}$) and water (H$_{2}$O).   Indeed, more than 90\% of the mantle consists of these two species at early times.   Both molecules form on the grain surface in competition with each other via the  reactions ${\rm grOH}+{\rm grCO}\rightarrow {\rm grCO_{2}}+{\rm grH}$ and ${\rm grOH}+{\rm grH}\rightarrow {\rm grH_{2}O}$.  Here gr indicates that species are on the grain surface. Atomic hydrogen comes from the gas phase where it is produced  via dissociation of H$_{2}$ by photons and cosmic rays (see Fig. \ref{all3-1}). About 50\% of the carbon monoxide is also accreted from the gas, but it is also produced on grain surfaces via the routes ${\rm grC}+{\rm grO_{2}}\rightarrow {\rm grCO}+{\rm grO}$ and ${\rm grC}+{\rm grOH}\rightarrow {\rm grCO}+{\rm grH}$. The common precursor of both CO$_{2}$ and H$_{2}$O, hydroxyl radical OH, is mainly formed from the atomic species accreting from the  gas.

\begin{figure*}\centering
\includegraphics[width=1.0\textwidth, angle=0]{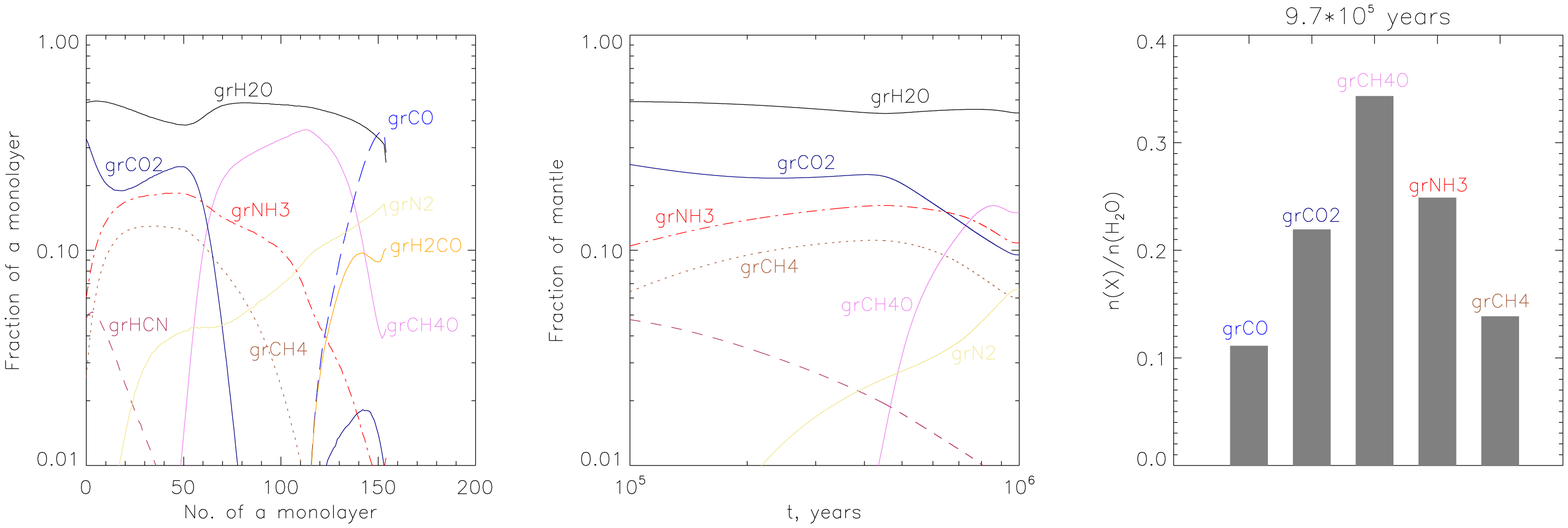}
\includegraphics[width=1.0\textwidth, angle=0]{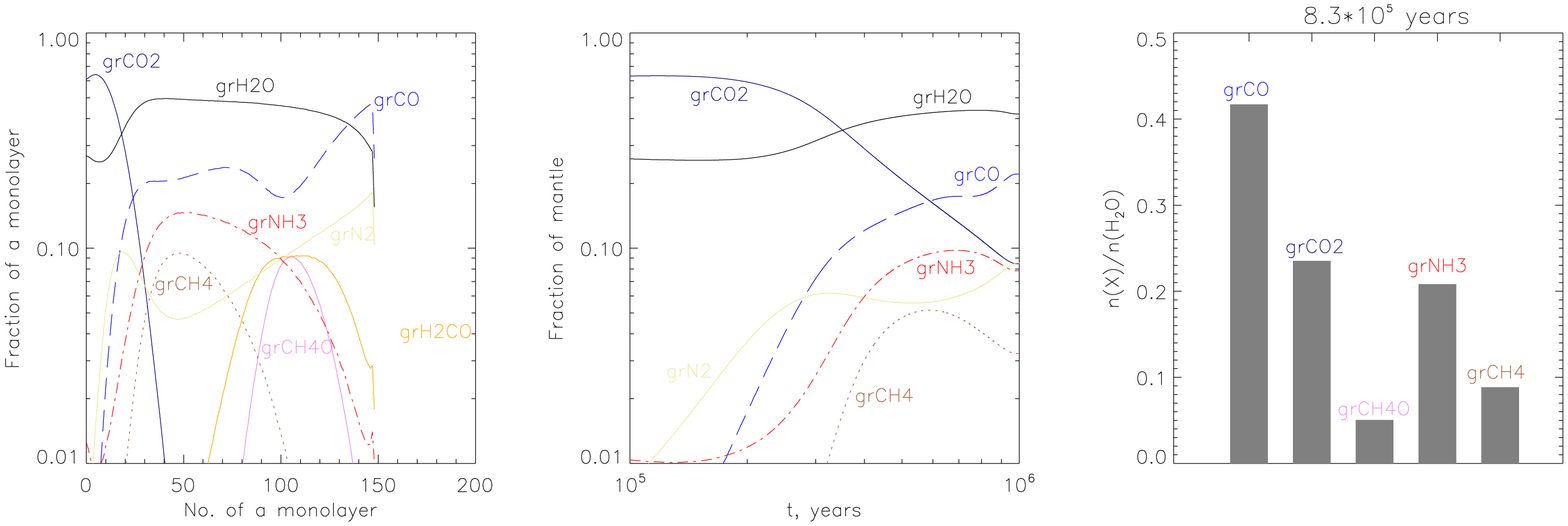}
\includegraphics[width=1.0\textwidth, angle=0]{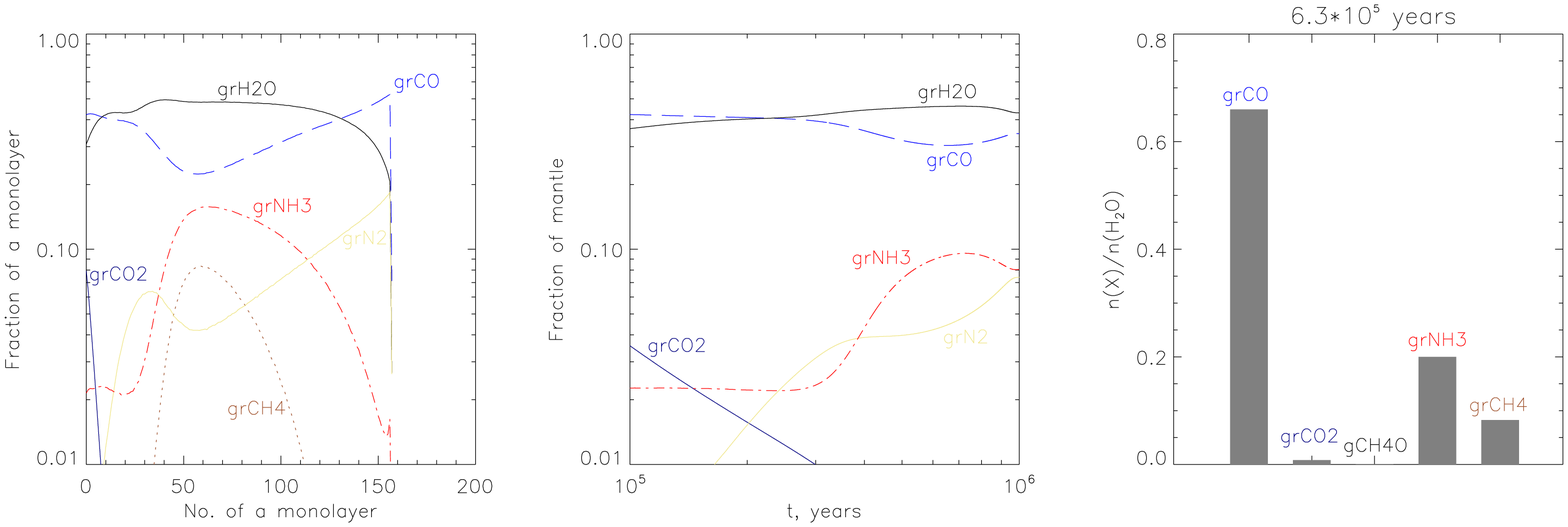}
\caption{Ice evolution for the MC-1 simulation  with $E_{\rm b}/E_{\rm D}$=0.3 (top row), $E_{\rm b}/E_{\rm D}$=0.5 (middle row), and $E_{\rm b}/E_{\rm D}$=0.77 (bottom row). In the left panels, in which fraction of a monolayer is plotted against individual monolayers, numbered from 0 at the surface, only species that contribute at least 5\% of at least one monolayer at any time are shown. In the middle panels, where fraction of the whole mantle is plotted vs time, only species that contribute at least 5\% of the entire bulk at any moment are shown. On histograms in the right column, abundances of the most abundant ice compounds with respect to solid water are shown at the time of minimum $F_{\rm fit}$.}\label{iceev14}
\end{figure*}

\begin{figure*}\centering
\includegraphics[width=1.0\textwidth, angle=0]{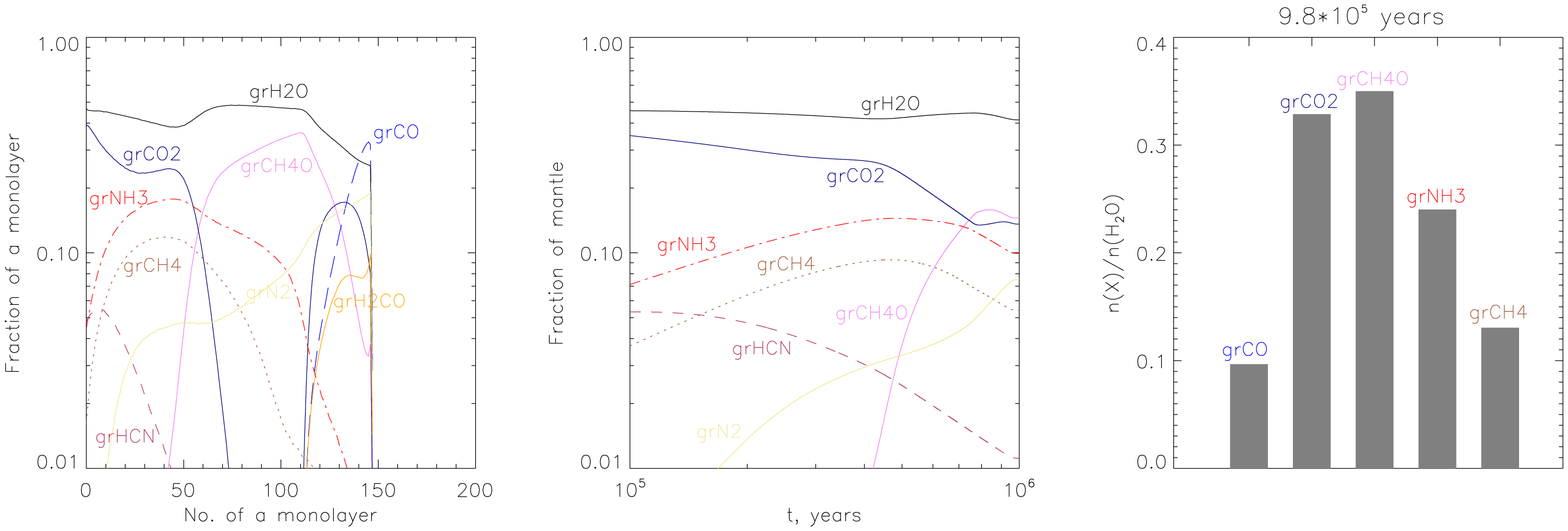}
\includegraphics[width=1.0\textwidth, angle=0]{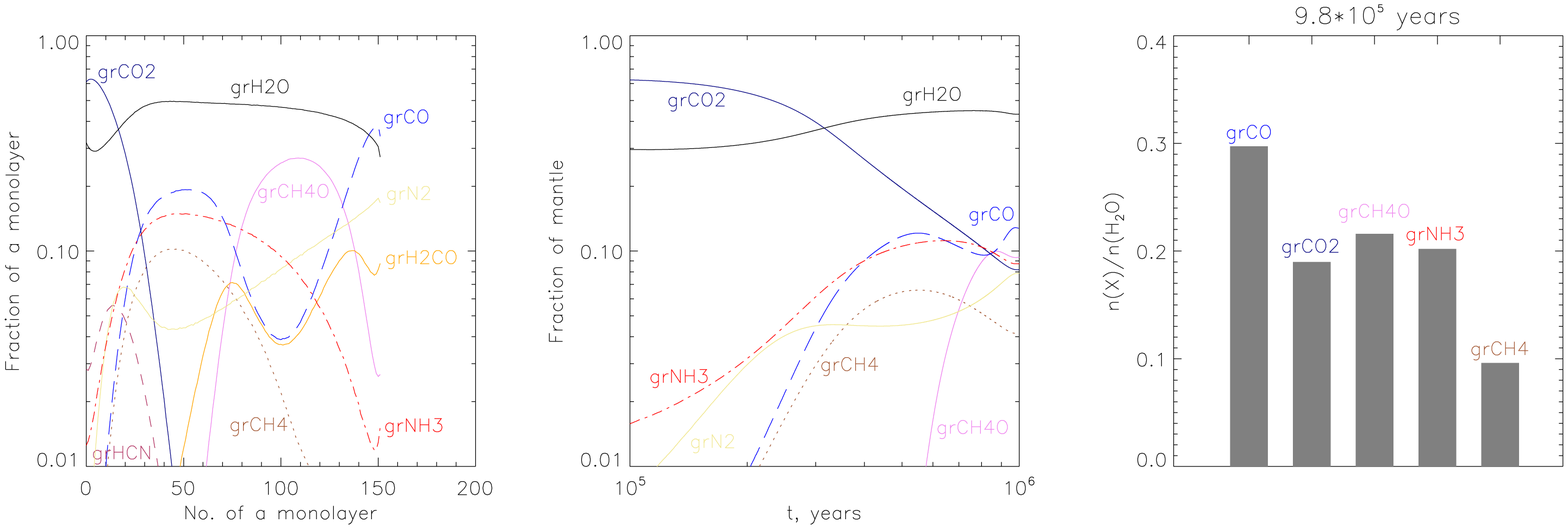}
\includegraphics[width=1.0\textwidth, angle=0]{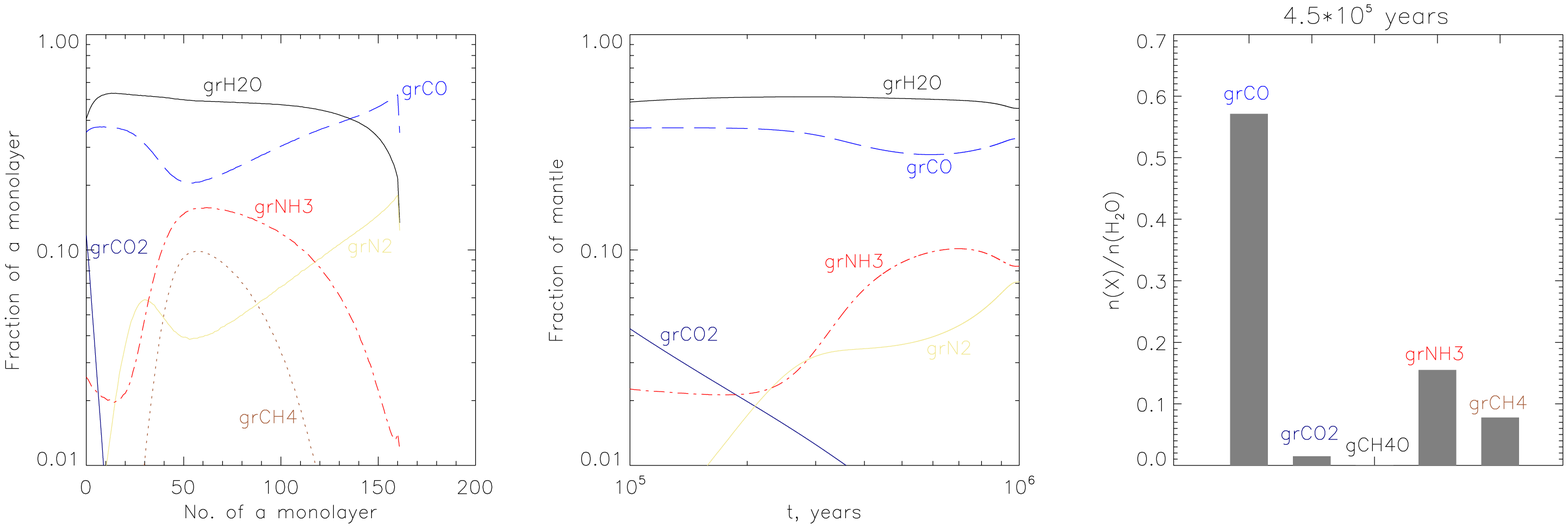}
\caption{The same as Figure~\ref{iceev14}, but for model MC-4.}
\label{iceev44}
\end{figure*}

In the multiphase models with $E_{\rm b}$/$E_{\rm D}= 0.3$, the competition between CO$_{2}$ and H$_{2}$O formation is won by water from the earliest times until the end of cold collapse phase. This is the most abundant compound in the ice since the first built monolayer in both Monte Carlo models MC-1 and MC-4, as can be seen in Figure~\ref{all9}, left column and also Figures~\ref{iceev14}, and~\ref{iceev44}, top row, middle panel. In the middle panels of these latter two figures, the molecular abundances are expressed in terms of fraction of the grain mantle, whereas in the left panels they are expressed in terms of fractions of individual monolayers.  In the RE model,  CO$_{2}$ is more abundant than H$_{2}$O.  We believe that this is due to stochastic effects in grain surface chemistry, which cannot be treated correctly in models based on rate equations \citep[][]{Vasyunin_ea09}.

In the models with $E_{\rm b}$/$E_{\rm D} =0.5$, the competition is won by CO$_{2}$ only in the early stage. In these models, the quantum tunneling as a source of atomic hydrogen mobility on the surface is switched off, and the majority of H atoms desorb back to the gas before they react with OH. The heavier molecules OH and CO are better bound to the surface, and therefore they can react at $T_{\rm dust}$=17-20~K, the relevant early stage temperature range.  Although an activation barrier of 80~K slows down this reaction, a combination of T$\sim$20~K, the relatively high mobility of CO at this temperature, and $E_{\rm b}/E_{\rm D}$=0.5 makes this reaction the major route of CO$_{2}$ formation.
For the model with $E_{\rm b}$/$E_{\rm D} =0.5$ at T=20~K, the mobility of OH is negligible: $\sim$10$^{-19}$ sites s$^{-1}$, while the mobility of CO is $\sim$1 site s$^{-1}$.

The ice composition in models with $E_{\rm b}$/$E_{\rm D}=0.77$ (corresponding to low surface mobility) differs from other models even at the early stage, as can be seen in Figure~\ref{all9}, right column, as well as the bottom row of Figures~\ref{iceev14} and \ref{iceev44}. The largest constituent is carbon monoxide CO  instead of water, while the  fraction of CO$_{2}$ in the entire mantle is always less than 10\%.  This dramatic difference is caused by the low mobility of surface species at $E_{\rm b}$/$E_{\rm D} =0.77$,  which makes the ${\rm grCO}+{\rm grOH}\rightarrow {\rm grCO_{2}}+{\rm grH}$ reaction inefficient in contrast to the model with $E_{\rm b}$/$E_{\rm D}$=0.5.  Thus, most of the CO remains in its pristine form in the ice mantle, where at the early stage it is formed in the ${\rm grC}+{\rm grO_{2}}\rightarrow {\rm grCO}+{\rm grO}$ and ${\rm grC}+{\rm grOH}\rightarrow {\rm grCO}+{\rm grH}$ reactions.

Two other major species in the ice mantle, ammonia NH$_{3}$ and methane CH$_{4}$,  have much smaller abundances than CO$_{2}$ and H$_{2}$O. Both NH$_{3}$ and CH$_{4}$ are formed  via similar chains of reactions: ${\rm grNH}\rightarrow {\rm grNH_{2}}\rightarrow {\rm grNH_{3}}$ and ${\rm grCH}\rightarrow {\rm grCH_{2}}\rightarrow {\rm grCH_{3}}\rightarrow {\rm grCH_{4}}$. Another abundant surface species at the early stage, molecular nitrogen (N$_{2}$) is mainly formed by the surface reactions ${\rm grN}+{\rm grN}\rightarrow {\rm grN_{2}}$ and ${\rm grNH}+{\rm grN}\rightarrow {\rm grN_{2}}+{\rm grH}$. Also, about 40\% of N$_{2}$ accretes from the gas phase.

Since visual extinction is low at early times ($A_{\rm V} \le 3$), the photodissociation of molecules on grain surfaces is quite active. At the same time, it does not strongly affect the  chemical composition of the ice, because the major molecules are involved in formation-destruction cycles such as ${\rm grH}+{\rm grOH}\rightarrow {\rm grH_{2}O}$, ${\rm grH_{2}O}+h\nu\rightarrow {\rm grH}+{\rm grOH}$. The net effect of such loops is mainly a somewhat delayed build-up of the mantle, which can be seen as a less steep slope in Figure~\ref{mantlegrowth} at early time.

\subsubsection{Intermediate collapse stage}
After $\sim 2 \times 10^{5}$ yr of collapse,  differences between models with various $E_{\rm b}$/$E_{\rm D}$ ratios as well as between the two-phase rate equation-based and multiphase Monte Carlo-based models become more pronounced.  The most noticeable transition is a sharp decrease in the fraction of carbon dioxide CO$_{2}$. In models MC-1 and MC-4 with $E_{\rm b}$/$E_{\rm D}=0.3$,  the 60 lowest monolayers of the mantle are rich in CO$_{2}$, and the sharp fall-off of its fraction happens at t$\sim 5 \times10^{5}$ yr, as can be seen in Figures~\ref{iceev14} and \ref{iceev44}.  For Monte Carlo models with $E_{\rm b}$/$E_{\rm D}=0.5$,  only $\sim 25$ monolayers are rich in CO$_{2}$, and the sharp fall-off of CO$_{2}$ fraction happens earlier, at t$ \sim 3 \times 10^{5}$ yr. The fall-off occurs as the reaction ${\rm grOH}+{\rm grCO}\rightarrow {\rm grCO_{2}}+{\rm grH}$ becomes inefficient, which occurs at 17~K if $E_{\rm b}$/$E_{\rm D}=0.5$ and 15~K if $E_{\rm b}$/$E_{\rm D}=0.3$. As the temperature falls, CO$_{2}$ formation loses the competition with H$_{2}$O formation. In models with $E_{\rm b}/E_{\rm D}$=0.3 it happens later than in models with $E_{\rm b}/E_{\rm D}$=0.5  because of the lower surface mobility of species at higher $E_{\rm b}/E_{\rm D}$.   From this point of view, models with $E_{\rm b}/E_{\rm D}$=0.77, where CO$_{2}$ formation is not efficient from the very beginning, can be considered as models where surface mobility of CO is too low to form CO$_{2}$ even at T$\sim$20~K. At the late stage, in models with $E_{\rm b}/E_{\rm D}$=0.3, the fraction of CO$_{2}$ increases again because the abundance of atomic hydrogen drops in the gas phase with increasing density, and the OH on the surface is not consumed in H$_{2}$O formation efficiently.

Interestingly, as can be seen on top left panel of the Figure~\ref{all9}, in the RE model with $E_{\rm b}/E_{\rm D}$=0.3,  CO$_{2}$ remains the most abundant fraction of the mantle until the end of the simulation. This arises from the fact that in the two-phase RE model the entire mantle is accessible for surface photodissociation processes, contrary to the multiphase MC-1 and MC-4 models, where only the upper monolayers can be photodissociated. The significant fraction of newly formed water is dissociated back to H and OH even at the late stage, because of cosmic ray-induced photodissociation. OH, in turn, can react with abundant CO and form CO$_{2}$, even though the OH+CO channel is relatively inefficient.  In multiphase models MC-1 and MC-4, the majority of water is buried within the thick mantle and is not accessible for photodissociation.This makes the final abundance of water higher in MC models than in the RE model and damps CO$_{2}$ formation at the late stage. Since CO$_{2}$ can be photodissociated too, in RE models with $E_{\rm b}/E_{\rm D}$=0.5 and $E_{\rm b}/E_{\rm D}$=0.77 one can see a late decrease in the CO$_{2}$ abundance. There, the mobility of CO is not enough to resist photodissociation, and the majority of CO$_{2}$ is converted back to CO. In the model with $E_{\rm b}/E_{\rm D}$=0.5, CO is ultimately converted to methanol, while in the model with $E_{\rm b}/E_{\rm D}$=0.77 it remains in its original form.

Another important chemical difference between the intermediate and early collapse stages in models MC-1 and MC-4 with $E{\rm b}$/$E_{\rm D}$=0.3 and $E{\rm b}$/$E_{\rm D}$=0.5 is the activation of the CO hydrogenation sequence ${\rm grCO}\rightarrow {\rm grHCO}\rightarrow {\rm grH_{2}CO}\rightarrow {\rm grH_{3}CO}\rightarrow {\rm grCH_{3}OH}$. This very important process is not efficient at early time because the temperature is too high for H atoms to be efficient at surface hydrogenation and because the high abundance of OH destroys formaldehyde in the reaction ${\rm grOH}+{\rm grH_{2}CO}\rightarrow {\rm grHCO}+{\rm grH_{2}O}$. Once the temperature decreases to $\le$15~K, the hydrogenation sequence becomes more effective. In models with $E_{\rm b}/E_{\rm D}$=0.3, CH$_{3}$OH becomes abundant at $\sim 5 \times 10^{5}$ yr, while its fraction in the newly formed mantle monolayers reaches $\sim$30\%. In models with $E_{\rm b}/E_{\rm D}$=0.5, on the other hand, the maximum abundance of methanol in these mantle monolayers is only $\sim$10\%, a number reached later, after $\sim 7 \times10^{5}$ yr of collapse. Formaldehyde (H$_{2}$CO), the precursor of methanol, is also formed in ice in significant amounts in the intermediate and late stages. In the RE model with $E_{\rm b}/E_{\rm D}$=0.3, the methanol fraction never becomes significant. Again, this is a manifestation of the two-phase approach, where methanol is efficiently photodissociated in the entire mantle to formaldehyde. About 25\% of the H$_{2}$CO, in turn, is photodissociated to CO and H$_{2}$. Finally, CO is converted to CO$_{2}$. In models with $E_{\rm b}/E_{\rm D}$=0.77 methanol is not produced in appreciable amounts.

\subsubsection{Late collapse stage}
\begin{figure*}\centering
\includegraphics[height=1.0\textwidth, angle=90]{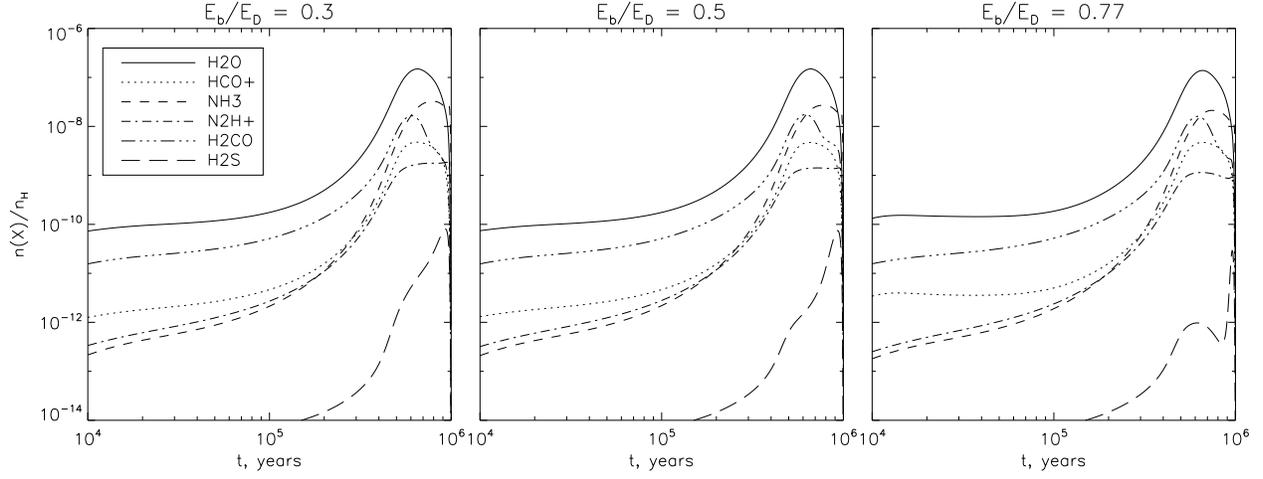}\vspace{5mm}\caption{The evolution of observationally important gas-phase species during the cold collapse phase.}\label{all3-2}
\end{figure*}

The late stage of collapse ($t \ge 7 \times 10^{5}$ yr) is characterized by a rapid increase in the gas density and visual extinction A$_{V}$, as can be seen in Figure~\ref{modpar}.  In the gas phase, the abundances of species first reach values typical for cold prestellar clouds (see Figure~\ref{all3-2} for $t \sim 7 \times10^{5}$ yr). Then, after $8-9 \times10^{5}$ yr,  gas-phase species rapidly deplete onto grains due to the high density and low temperature. Among these species is the most abundant gas phase molecule, carbon monoxide. The composition of ice monolayers formed during this period is characterized by an increasing fraction of CO, as seen in Figures~\ref{iceev14} and \ref{iceev44}. Correspondingly, the fraction of other ice constituents decreases in the outer monolayers. In all Monte Carlo models, CO becomes the most abundant ice species  in the outer 10-30 monolayers, depending on $E_{\rm b}/E_{\rm D}$.

The analysis above shows that one can delineate three chemically different phases of the ice mantle build-up during the contraction of a cold core in multiphase models with $E_{\rm b}$/$E_{\rm D}$=0.3 and $E_{\rm b}$/$E_{\rm D}$=0.5. At $t \le 2 \times10^{5}$ yr, the CO$_{2}$-rich fraction of the ice mantle is formed, and is intimately mixed with water, which is  formed at the same time. Thus, we can attribute this fraction of the ice to the polar CO$_{2}$:H$_{2}$O component of ice inferred from observations \citep[][]{Tielens_ea91, Oeberg_ea11}. Between $ 2 \times 10^{5}$ yr and $8 \times 10^{5}$ yr, an H$_{2}$O-rich ice is formed. Other major ice constituents at this time are CO, ammonia NH$_{3}$, formaldehyde H$_{2}$CO, and methanol CH$_{3}$OH. The third phase of ice is  formed during the late stage of collapse between $ 8 \times 10^{5}$ yr and 10$^{6}$ yr. This period is characterized by the massive freeze-out of CO from the gas phase onto the grain surface. Thus, the outer layers of the ice mantle are dominated by solid CO and may be associated with the  inferred apolar component of interstellar ices \citep[][]{Tielens_ea91, Oeberg_ea11}.  This picture, generally consistent with observations, is not  clearly pronounced  in the two-phase RE models considered in this study.  Instead,  there are no inert ice monolayers the composition of which reflects the history of changing physical conditions during the ice formation. Rather, the entire mantle is available for chemical reactions all the time, which leads to a mixed ice composition.  In models with $E_{\rm b}/E_{\rm D}$=0.77 the first CO$_{2}$-rich phase of the ice mantle is never produced. This failure to reproduce observational inference argues against an $E_{\rm b}/E_{\rm D}$ ratio equal to 0.77.

\subsubsection{Comparison with observations}
It is interesting to compare the modeling results with observational data of cold ices. To do this,  we took the median ice composition from Table~2 of \citet[][]{Oeberg_ea11}, which contains sources involving low- and high-mass protostars as well as background stars. These data are based on the critical compilation of observational data obtained with the {\it Spitzer} and {\it ISO} infrared space telescopes, and recommended by the authors as a reference for ice formation models. For comparison with our simulations, we used the median ice composition for low-mass protostars because our temperature profile during collapse corresponds to the early stage of cloud contraction. Thus, there is no protostellar heating phase,  which is probably responsible for the differences in ice composition between low-mass and high-mass protostars in later stages of evolution \citep[][]{Oeberg_ea11}. The protostellar heating phase can rather be attributed to the warm-up stage of the model, where temperature increases from 10~K to 200~K occur. One should bear in mind that the observed ice composition in each category is obtained from the analysis of infrared absorption spectra, which are usually averaged over regions with different physical conditions. In our model, we consider just one parcel of a gas-grain mixture with homogeneous (although time-dependent) physical conditions. Therefore, some discrepancies between observed and modeled ice compositions are natural.

\begin{figure*}\centering
\includegraphics[width=01.0\textwidth]{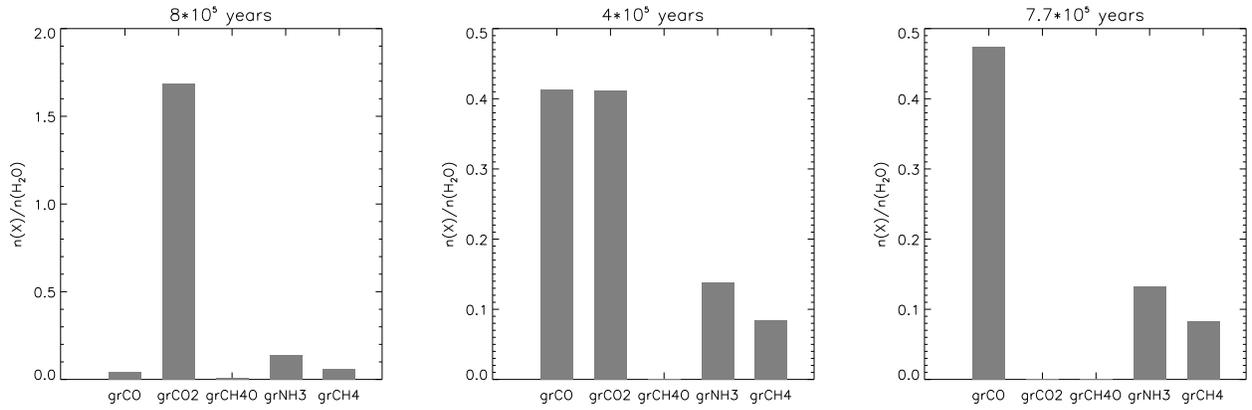}\vspace{5mm}\caption{Best fit ice composition in RE models. Left panel: ${\rm E_{b}}/{\rm E_{D}}$=0.3, middle panel: ${\rm E_{b}}/{\rm E_{D}}$=0.5, right panel: ${\rm E_{b}}/{\rm E_{D}}$=0.77.}\label{rebestfit}
\end{figure*}

Our time-dependent values for the fractional abundances of major ice species per monolayer and per mantle  are shown in Figures~\ref{iceev14} and \ref{iceev44} for the MC-1 and MC-4 simulations respectively.  We compare our results for overall abundances per mantle with observed median abundances as a function of time to determine the optimum agreement time  $t_{\rm comp}$, when  the sum of squares of the residual differences is minimal:
\begin{equation}
t_{\rm comp}:F_{\rm fit}=\sum_{i=1}^{N}(X_{\rm i}^{obs}-X_{\rm i}^{model}(t))^2=min.
\end{equation}
Here, $t_{\rm comp}$ is the time of  best agreement, $X_{\rm i}^{obs}(t)$ is the observational abundance of species $i$ with respect to the abundance of solid water, while $X_{\rm i}^{model}(t)$ is the modeled abundance of the same species at time $t$. The smaller the value of $F_{\rm fit}$, the closer the modeled ice composition to the observed composition.  Histograms of the best fit to the ice composition for MC models are shown in the right column of Figures \ref{iceev14} and \ref{iceev44}, while histograms of the best fit to the observed ice composition for RE models are shown in Figure~\ref{rebestfit}.

The optimum time and ice composition corresponding to the minimum values of $F_{\rm fit}$ for different models are summarized in Table~\ref{ice_comp_table} along with the observed median values for low-mass protostars, with the water ice abundance fixed at 100. $F_{\rm fit}$ changes with time smoothly, so one should consider the best fit time as a rough estimation, not a precise value. One can see that reasonable ice compositions are produced by MC and RE models with $E_{\rm b}/E_{\rm D}$=0.5. Models with $E_{\rm b}/E_{\rm D}$=0.3 underproduce solid CO.   At the same time, the calculated amount of CO$_{2}$ is very high in the RE model (170\% of that of water), while in the MC models CH$_{3}$OH is significantly overproduced. Specifically, in the MC $E_{\rm b}/E_{\rm D}$=0.3 models, the optimum CO$_{2}$ abundance lies between 22\% and 32\%, while the methanol abundance  is $\sim$34\%--35\%. In addition, the methane abundance in these models is 2--3 times higher than the observed value.
Models with $E_{\rm b}/E_{\rm D}$=0.77, on the contrary, overproduce the solid CO fraction and greatly underproduce CO$_{2}$ and methanol. Specifically, the CO fraction varies from $\sim$50\% in the RE model to $\sim$50\%-70\% in the MC-1 and MC-4 models. At the same time, the CO$_{2}$ and CH$_{3}$OH fractions are calculated to lie close to zero. The calculated methane abundance in models with $E_{\rm b}/E_{\rm D}$=0.77 is in reasonable agreement with observation, at only 1.5--2.0 times higher than the observed value.

Of the three  models that yield very low values of $F_{\rm fit}$, the multiphase Monte Carlo models with $E_{\rm b}/E_{\rm D}$=0.5 produce mantle compositions closer to what is observed. Of these two, Model MC-1 produces the better fit as indicated by the lower value of $F_{\rm fit}$=0.03 for MC-1 vs. $F_{\rm fit}$=0.07 for MC-4.   For MC-1, the modeled values are closer to median detected values, and deviate from them by a factor of 1.5-2.0, except for NH$_{3}$.  Model MC-4, on the other hand, overproduces methanol and underproduces CO$_{2}$. Although the RE model with $E_{\rm b}/E_{\rm D}$=0.5 possesses the minimum value of $F_{\rm fit}$=0.02,  it should be remembered that $F_{\rm fit}$ characterizes the difference between modeled and observed abundances of ice constituents with respect to solid water. In the RE model with $E_{\rm b}/E_{\rm D}$=0.5 an optimum ice composition is produced at an optimum time of $\sim 4 \times 10^{5}$ yr, when the conditions have not yet reached high density and A$_{V}$, and the abundances of ice constituents with respect to hydrogen number density are  at least an order of magnitude too low to explain the observed values.

Let us now compare the modeled abundances of selected gas-phase species shown in Figure~\ref{all3-2} with observational values. These abundances reach the peak values at $\sim$6$\times$10$^{5}$~yr when the gas density reaches $\sim$10$^{4}$~cm$^{-3}$. Later, due to high density, the gas-phase species are depleted on grains rapidly, and their abundances get very low. Our modeled peak abundances of gas-phase species are in reasonable agreement with observational values for the dark cloud core TMC-1~(CP),  as seen in  Table~\ref{gas_phase_species}.

\subsection{Warm-up phase of a hot core}
We now discuss the chemistry during the warm-up stage leading to the formation of a hot core, using both two-phase and multiphase models.  As discussed in the previous section, we ran the RE, MC-1 and MC-4 models, each with the three different values of $E_{\rm b}/E_{\rm D}$.

\begin{figure*}\centering
\includegraphics[height=1.0\textwidth, angle=90]{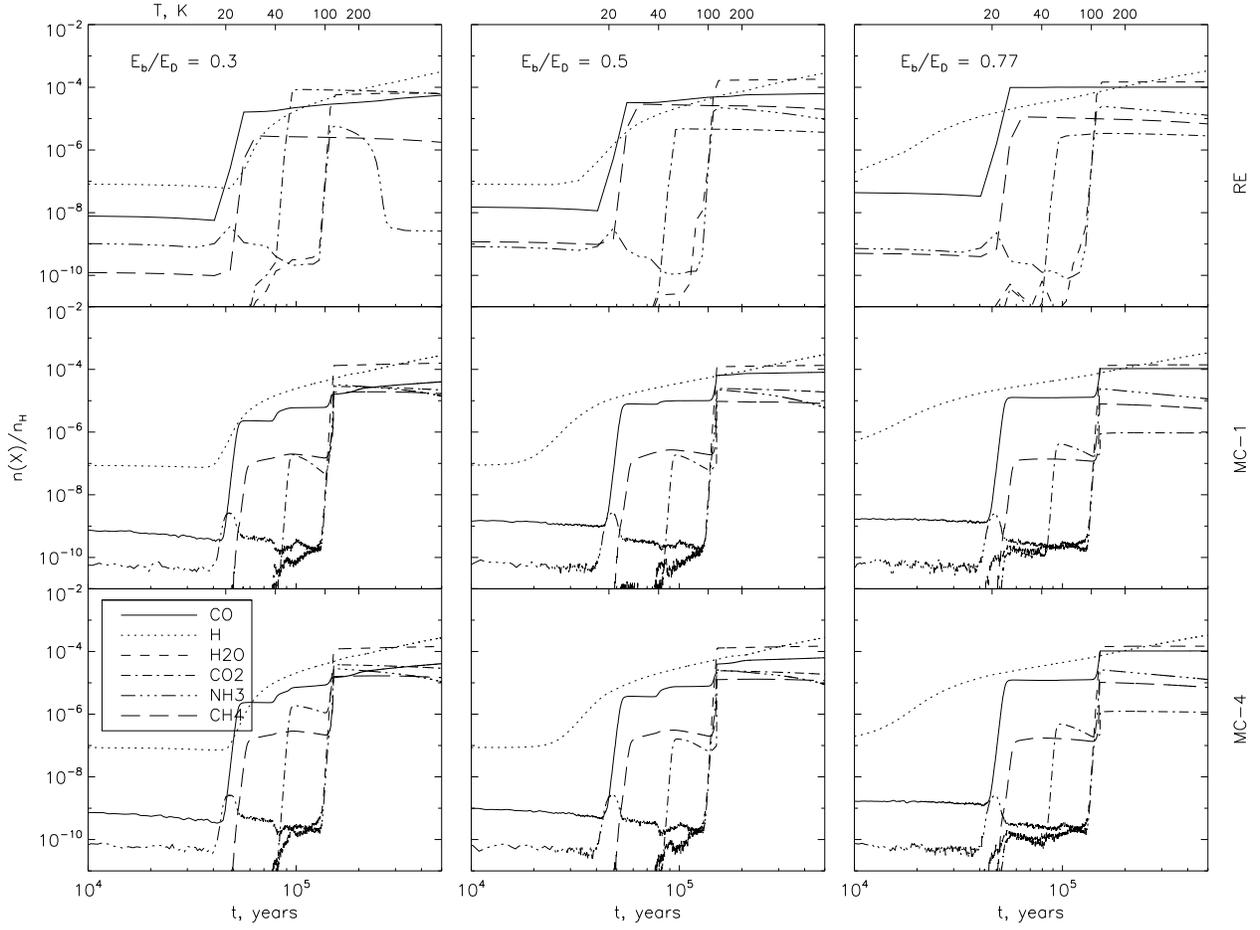}\vspace{5mm}\caption{Abundances of major gas-phase species during the warm-up phase.}\label{f11}
\end{figure*}

\begin{figure*}\centering
\includegraphics[height=1.0\textwidth, angle=90]{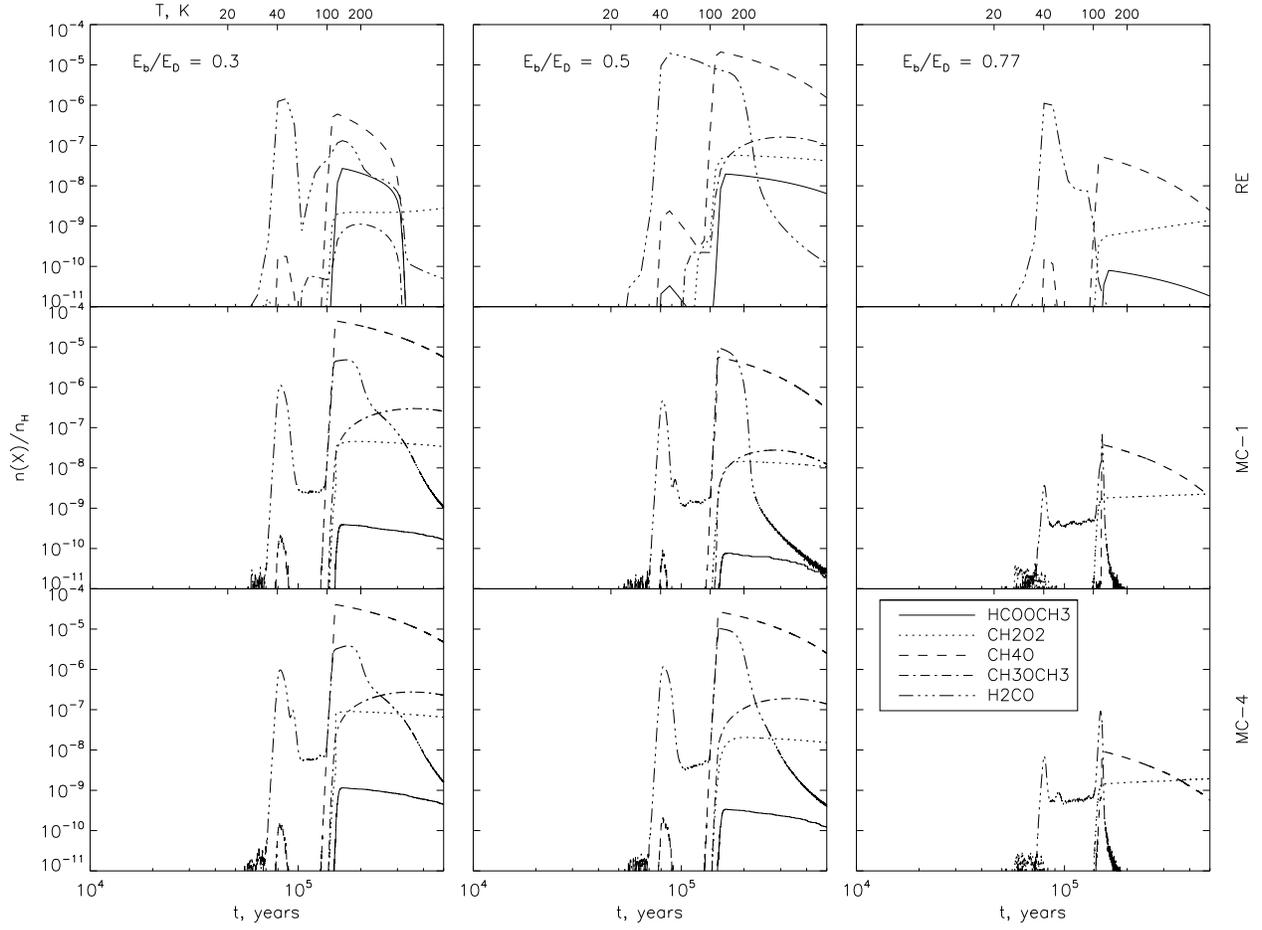}\vspace{5mm}\caption{Abundances of organic gas-phase species during the warm-up phase.}\label{f12}
\end{figure*}

The chemical evolution during the warm-up phase is presented in Figures \ref{f11}, \ref{f12}, and \ref{surfspecwarmup}. In our model, the warm-up stage begins after the cold collapse stage is finished; i.e., after 10$^{6}$~yr of evolution. In  Figures \ref{f11}, \ref{f12}, and \ref{surfspecwarmup} the warm-up time starts from zero, but the zero point corresponds to the end of the cold collapse phase. The warm-up from 10~K to 200~K occurs from 0~yr to 2$\times$10$^{5}$~yr. Figure~\ref{f11} shows the evolution of gas-phase species  that are also main constituents of ice mantles during the early warm-up stage, having  accumulated on grain surfaces during the cold collapse phase.  During the warm-up, they are eventually ejected into the gas phase by thermal and non-thermal mechanisms. In Figure~\ref{f12}, the evolution of gas-phase oxygen-containing organic species formed via a combination of gas-phase and grain-surface chemistry during the warm-up phase is shown. Figure~\ref{surfspecwarmup} exhibits the evolution of their grain surface counterparts. The noise in low abundance seen in the figures is an intrinsic feature of Monte Carlo simulations,  which deal with whole numbers of species. In our model, an absolute abundance of unity corresponds to an abundance relative to the total number of hydrogen nuclei of $\sim$10$^{-12}$.

\subsubsection{Abundant species}
The top row of  Figure~\ref{f11} represents the results of the RE models. Our two-phase RE  model with $E_{\rm b}/E_{\rm D}$=0.5 essentially reproduces the results obtained by \cite{GarrodHerbst06}, who used a similar approach.   All species  except atomic hydrogen are desorbed  into the gas phase from the ice mantle at their sublimation temperatures.  In the RE model, species evaporate independently of each other, and evaporation happens in one step, after which abundances of these rather stable species stay nearly constant until the end of the simulation.  The abundances of the gas-phase species, once sublimation has occurred, are very similar to the abundances in the ice before sublimation occurs. The exception is ammonia,  which drops by 3 orders of magnitude in less than $ 2 \times 10^{5}$ yr after sublimation in model RE with $E_{\rm b}/E_{\rm D}$=0.3.

The middle and bottom rows of  Figure~\ref{f11} show the results obtained with the multiphase Monte Carlo models MC-1 and MC-4. The main difference from the RE model is the two-step picture of the sublimation of volatile species discussed earlier in this paper. The first sublimation step occurs when sublimation temperature of the volatile species is reached, at which time only the species from four chemically active monolayers desorb. The second step occurs at the sublimation temperature of water ice, which releases all trapped volatile species  as well. The two-step phenomenon is a new phenomenon in warm-up chemistry (if not in the laboratory), completely missing in two-phase models. Interestingly, after the first sublimation step, CO$_{2}$ in the gas phase has a similar abundance in models with $E_{\rm b}/E_{\rm D}$=0.77 despite the fact that there was no CO$_{2}$ on the ice surface at the end of cold collapse phase. This CO$_{2}$ is produced initially on the ice surface during the early warm-up when the dust temperature reaches 20~K. This early CO$_{2}$ may  contribute to the formation of a segregated layer of pure CO$_{2}$ observed towards protostars, and usually considered as  evidence of moderate ice heating \citep[][]{Boogert_ea08, Pontoppidan_ea08}.

\subsubsection{Complex organic species}
\begin{figure*}\centering
\includegraphics[height=1.0\textwidth, angle=90]{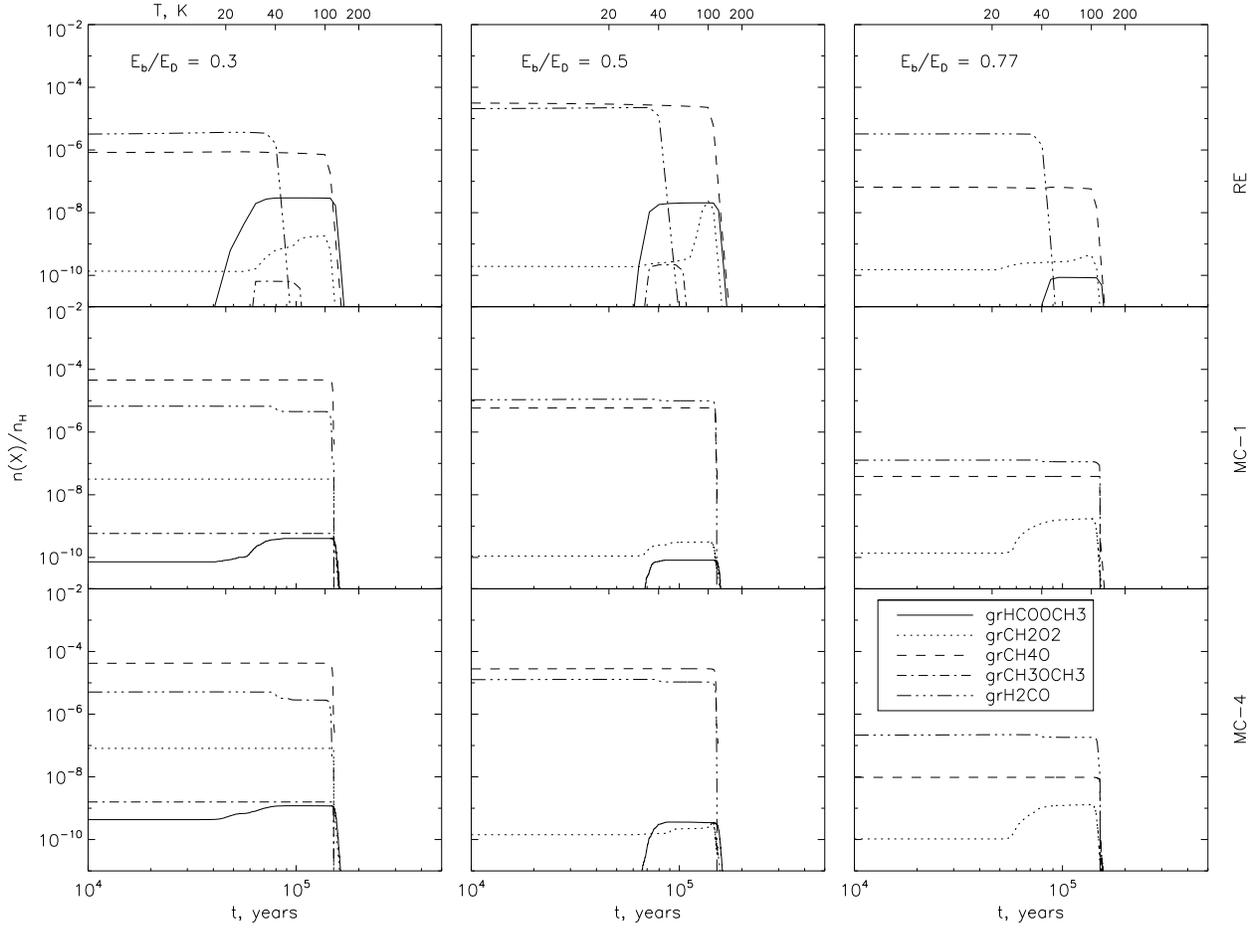}\vspace{5mm}\caption{Abundances of organic species on the grain surface during the warm-up phase.}\label{surfspecwarmup}
\end{figure*}

In Figure~\ref{f12}, the gas-phase abundances of organic molecules produced mainly on warming ice surfaces followed by sublimation are presented as functions of time and rising temperatures. Methanol and formaldehyde, as precursors of these species, are also shown. Figure~\ref{surfspecwarmup} shows the time dependence of the abundances of these molecules in their solid phase, which are formed by radical-radical recombination during warm-up. The formation of these species in the framework of two-phase models has been discussed in detail by
\citet{GarrodHerbst06, Garrod_ea08}, and \citet{Laas_ea11}. It is  interesting to see how the new multilayer approach to surface chemistry affects the previously proposed mechanisms.

According to the analysis performed by \cite{GarrodHerbst06}, methyl formate HCOOCH$_{3}$ is formed primarily on grain surfaces via the reaction
\begin{equation}\label{HCOOCH3formsurf}
{\rm grHCO}+{\rm grCH_{3}O}\rightarrow {\rm grHCOOCH_{3}},
\end{equation}
assisted by a gas-phase reaction:
\begin{equation}\label{HCOOCH3formgas}
{\rm H_{2}COH^{+}}+{\rm H_{2}CO}\rightarrow {\rm H_{2}COHOCH_{2}^{+}+h\nu}.
\end{equation}
In this work, we make no distinction between the {\it cis} and {\it trans} forms of methyl formate, as discussed by \citet{Laas_ea11}.
Although formic acid HCOOH and dimethyl ether CH$_{3}$OCH$_{3}$ also have surface formation routes:
\begin{equation}
{\rm grHCO}+{\rm grOH}\rightarrow {\rm grHCOOH},
\end{equation}
and
\begin{equation}\label{CH3OCH3formsurf}
{\rm grCH_{3}}+{\rm grCH_{3}O}\rightarrow {\rm grCH_{3}OCH_{3}},
\end{equation}
the majority of these two species are produced by gas-phase routes. During the early warm-up, HCOOH is formed in the gas phase via the dissociative recombination of HCOOH$_{2}^{+}$. The sharp rise in the HCOOH gas phase abundance during the late warm-up is caused by the gas-phase reaction
\begin{equation}\label{HCOOHform}
{\rm OH}+{\rm H_{2}CO}\rightarrow {\rm HCOOH}+{\rm H},
\end{equation}
which becomes effective after the hydroxyl radical desorbs from grains. For dimethyl ether, the main formation route is a gas-phase reaction of methanol and protonated methanol:
\begin{equation}\label{CH3OCH3form}
{\rm CH_{3}OH}+{\rm CH_{3}OH_{2}^{+}}\rightarrow {\rm CH_{3}OCH_{4}^{+}}+{\rm H_{2}O},
\end{equation}
followed by dissociative recombination.  This reaction becomes important when methanol evaporates from the grain mantle.

All models with $E_{\rm b}/E_{\rm D}$=0.77 are characterized by a low amount of formaldehyde and methanol in cold grain mantles as seen in panels of the right column of Figure~\ref{surfspecwarmup}.  As a result, the peak abundances of these species in the gas phase after sublimation during the warm-up phase are 1--4 orders of magnitude lower than in models with $E_{\rm b}/E_{\rm D}$=0.3 and $E_{\rm b}/E_{\rm D}$=0.5. Since the gas-phase route of formation for dimethyl ether (see eq.~\ref{CH3OCH3form}) becomes inefficient, the peak gas-phase fractional abundance of CH$_{3}$OCH$_{3}$ in the warm-up phase drops by 5 orders of magnitude to less than 10$^{-11}$. In the case of HCOOH,  the route via eq.~\ref{HCOOHform} is assisted by another gas-phase route to produce HCOOH via the recombination of CH$_{3}$O$_{2}^{+}$. This ion  is produced following the evaporation of water via the radiative association reaction
\begin{equation}\label{HCOOHform2}
{\rm HCO^{+}}+{\rm H_{2}O}\rightarrow {\rm CH_{3}O_{2}^{+} + h\nu}
\end{equation}
although reactions of this sort have little experimental information concerning them and the route must be regarded as tentative at best. If this route does occur, as in our models, the abundance of HCOOH is not significantly affected. Thus, the abundance of HCOOH when the temperature reaches 100~K in both MC and RE models with $E_{\rm b}/E_{\rm D}$=0.77 is lower by $\sim$2 orders of magnitude in comparison with models with $E_{\rm b}/E_{\rm D}$=0.5. Formation of the surface counterparts of {formic acid and dimethyl ether (see Figure~\ref{surfspecwarmup}) occurs mainly by the accretion of species formed in the gas phase, and marginally by surface formation routes, which are inefficient due to low mobility of radicals in models with $E_{\rm b}/E_{\rm D}$=0.77. The surface formation route of HCOOCH$_{3}$ (eq.~\ref{HCOOCH3formsurf}) is inefficient in models MC-1 and MC-4, too.  The low abundance of H$_{2}$CO in the gas after evaporation inhibits the gas-phase formation of methyl formate via reaction (\ref{HCOOCH3formgas}).  The inefficiency of both gas phase routes and surface routes of formation leads to the complete absence of methyl formate in models MC-1 and MC-4 and allows only a small amount of the molecule in the RE model through the surface production-sublimation route.  Below we discuss why surface formation of methyl formate via reaction (\ref{HCOOCH3formsurf})} is more efficient in two-phase models than in multiphase models.

As can be seen in Figures~\ref{f12} and \ref{surfspecwarmup}, models MC-1 and MC-4 with $E_{\rm b}/E_{\rm D}$=0.3 and $E_{\rm b}/E_{\rm D}$=0.5 exhibit qualitatively similar results to each other. There are two important differences from the results obtained with the two-phase RE model with $E_{\rm b}/E_{\rm D}$=0.5, which can be regarded as a reference. First, the behavior of formaldehyde and CO in the gas phase is more complex in multiphase models. While in the reference RE model, H$_{2}$CO has a single desorption peak near the temperature of sublimation followed by a gradual destruction via gas-phase chemical reactions, in multiphase models there are two peaks. The first peak arises near $7 \times10^{4}$ yr and corresponds to the desorption of formaldehyde from the top four monolayers of the grain mantle. This peak occurs with a fractional abundance of $\sim$10$^{-6}$. During the next $ 2 \times 10^{4}$ yr, the H$_{2}$CO abundance drops to 10$^{-9}$, before the second peak occurs. This peak arises at $\sim 1.2 \times 10^{5}$ yr and corresponds to the desorption of refractory water ice, which also sweeps out the remaining volatile materials. After the second peak, formaldehyde is gradually consumed by ion-molecule reactions with protonating ions such as  HCO$^{+}$ and the neutral-neutral reaction with NH$_{2}$, which  attains a fractional abundance of  $\sim10^{-6}$ after the sublimation of its precursor, ammonia, from the ice mantle.
The CO behavior is  similar, with the same two peaks in multiphase models, one of which is missing in two-phase models.

The second important feature of the results obtained with multiphase models is the reduced abundance of gas-phase and grain surface methyl formate in comparison to the reference RE model. The peak abundances of both gaseous and ice methyl formate in our multiphase models lie up to a factor of hundred below their values in the reference RE model. In model MC-4 with $E_{\rm b}/E_{\rm D}$=0.5, the peak abundance of methyl formate is four times higher than in model MC-1. Thus, the abundance of methyl formate correlates strongly with the rate of surface photodissociation. The highest rate is in RE models, where all molecules in all $\sim$100 monolayers of mantle are available for surface photodissociation. In MC-4 models, only molecules in four chemically active monolayers can be photodissociated. Finally, in MC-1,  the  rate of surface photodissociation is effectively reduced by a factor of 4 in comparison with the MC-4 model. The rate of surface photodissociation is  important for the abundance of HCOOCH$_{3}$ because the radical precursors (OH, HCO) of surface methyl formate  are produced by photodissociation.

The efficiency of the grain-surface formation route of dimethyl ether (eq.~\ref{CH3OCH3formsurf}) also depends on the number of mantle monolayers available for photodissociation, because the methyl radical CH$_{3}$ is mainly produced by surface photodissociation of methanol. Nevertheless, the final gas-phase abundance of dimethyl ether is similar in two-phase and multiphase models, as the dominant formation route of CH$_{3}$OCH$_{3}$ is a gas-phase process (reaction~\ref{CH3OCH3form}).

\section{Discussion}\label{discussion}
In this study, we combined an advanced chemical model with a rather basic representation of the evolution of a core from its prestellar to protostellar phase. It is interesting to compare our results with  previous theoretical studies and the results of observations.

\subsection{Cold collapse phase}
Our model of the cold collapse phase is quite similar to the model described in Section 3.3 of \cite{GarrodPauly11}. In that study, the authors employed a modified version of a three-phase gas-surface-inert bulk chemical model initially proposed by \cite{HasegawaHerbst93}. The variations of physical conditions during the cold collapse are quite similar in \cite{GarrodPauly11} and our model. We use the same density and extinction profile, but a somewhat different profile of temperature. Some differences in surface chemistry are more significant. While \cite{GarrodPauly11} utilized a so-called reaction-diffusion competition for surface reactions with activation barriers \citep[][]{HerbstMillar08}, we stick to a more traditional approach \citep{Hasegawa_ea92}, in which such competition is not considered. This difference leads to the fact that surface reactions with barriers such as CO$_{2}$ formation via the route ${\rm grCO}+{\rm grOH}\rightarrow {\rm grCO_{2}}+{\rm grH}$ proceed more efficiently in \cite{GarrodPauly11} than in our model. Another important difference is the mechanism of oxygen hydrogenation on top of a buried CO molecule implemented in \cite{GarrodPauly11}. We do not have this mechanism in our model, so the CO$_{2}$ production efficiency at low temperatures in our model is further reduced.  Nevertheless, the ice composition and structure in our stochastic multiphase models with $E_{\rm b}/E_{\rm D}$=0.3 and $E_{\rm b}/E_{\rm D}$=0.5 are quite similar to those obtained in the collapse model of \cite{GarrodPauly11}. Interestingly, while \cite{GarrodPauly11} use $E_{\rm b}/E_{\rm D}$=0.3 in their collapse model, our models with the closest ice composition, MC-1, has $E_{\rm b}/E_{\rm D}$=0.5. This is because the temperature during collapse in our model is higher by a few degrees  than in \cite{GarrodPauly11}. This leads to higher mobilities of species on a grain surface in our model.

The main conclusion from the comparison of our model with that of  \cite{GarrodPauly11} for the cold collapse phase is that the observed ice structure and composition can be qualitatively reproduced with an approach in which cold collapse occurs with a drop in temperature combined with a stochastic multilayer model or an approximately stochastic three-phase treatment \citep{GarrodPauly11}, each of which protects already formed ice monolayers from further chemical processing.  The adopted details of grain surface chemistry such as the new CO$_{2}$ formation mechanism via the surface hydrogenation of oxygen on top of a buried CO molecule and reaction-diffusion competition are less important. The traditional treatment of the surface reaction ${\rm grCO}+{\rm grOH}\rightarrow {\rm grCO_{2}}+{\rm H}$ alone is sufficient to produce observed amounts of carbon dioxide on dust grains.

Another control parameter of the grain surface chemistry --- the diffusion/desorption energy ratio --- is of crucial importance. As was shown above, the cold ice composition varies strongly in models with different adopted $E_{\rm b}/E_{\rm D}$ values. Unfortunately, this parameter is known poorly, because it is difficult to measure the diffusion barrier in laboratory experiments.   It is tempting to conclude that comparison of our modeling results with observations of ices towards low-mass protostellar cores support a diffusion/desorption energy ratio of 0.5. However, one should be careful here. First, our model as well as that of  \cite{GarrodPauly11} uses for comparison a median ice composition towards low-mass protostars from the infrared surveys summarized in \cite{Oeberg_ea11}. While the average values are similar to those obtained with our model MC-1 ($E_{\rm b}/E_{\rm D}$=0.5), the ice composition towards individual sources exhibits significant variations. These variations probably reflect the different evolutionary stages of individual sources and different level of energetic processing of ices \citep[e.g.,][]{Gibb_ea00}. For example, there is evidence that CO$_{2}$ might be more abundant than CO in evolved protostars such as W33a \citep[][]{Gibb_ea00}. Also, evolved high mass protostars such as W33a and GL~7009S show a significantly higher than average abundance of methanol.  The recent study of \cite{Aikawa_ea12} also puts upper limits on the CH$_{3}$OH abundance in the low-mass protostars L1527 and IRAS04302 to be 26\% and 42\% respectively with respect to water; these values are significantly higher than the median abundance.  Such an ice composition is closer to our models MC-1 and MC-4 with $E_{\rm b}/E_{\rm D}$=0.3. Second, the ice composition depends on the ice thermal history during the cold collapse phase. In this study, a very simple temperature profile is utilized, which may differ from the thermal history of real ices.  We did not explore this because of the  huge demand of Monte Carlo models for CPU time. These facts indicate that comparison of our modeling results with observations is not sufficient to put strong constraints on the $E_{\rm b}/E_{\rm D}$ ratio. We can only claim that models with $E_{\rm b}/E_{\rm D}$=0.77 are eliminated by comparison with observations.

Another interesting result crucially dependent on the $E_{\rm b}/E_{\rm D}$ ratio in our study is the amount of reactive radicals trapped inside the bulk ice.   \citet[][]{Taquet_ea12} found in their model that a multilayer approach with ice porosity taken into account leads to relatively high abundances of reactive radicals in the bulk  ice. In their reference model (constant density of 10$^{5}$~cm$^{-3}$, constant temperature of 15~K, $E_{\rm b}/E_{\rm D}$=0.65), the fractional abundances of grOH, grHCO and grCH$_{3}$O trapped in the ice reach $\sim$5$\times$10$^{-6}$, $\sim$10$^{-8}$ and $\sim$10$^{-9}$, respectively. We observe similarly high abundances of these radicals only in our multiphase models with $E_{\rm b}/E_{\rm D}$=0.77. Thus, in addition to the multilayer approach,  which protects radicals by burying them in the bulk ice, a low mobility of surface species is essential, too; reactive radicals should be buried by accreting species before they react with other species on the surface. \citet[][]{Taquet_ea12} propose that abundant radicals are synthesized in the cold ice without photolysis, and then trapped, so that they may serve as precursors of complex organic species during the warm-up phase. However, as one can see at the right column of Fig.~\ref{f12} and Fig.~\ref{surfspecwarmup}, in our models, relatively high abundances of reactive radicals in the ice do not lead to the efficient formation of complex organic species in models with high $E_{\rm b}/E_{\rm D}$=0.77. During the warm-up, radicals are released at the temperature of sublimation of water ice, which is close to 100~K, and quickly evaporate where they are consumed in ion-molecular and photodissociation reactions.

As shown in \cite{Vasyunin_ea09} and \cite{Garrod_ea09}, models of surface chemistry with $E_{\rm b}/E_{\rm D}$=0.3 exhibit significant stochastic effects, while models with $E_{\rm b}/E_{\rm D}$=0.77 do not. Although the importance of stochastic effects in models with $E_{\rm b}/E_{\rm D}$=0.5 was not studied in detail here, one may assume that in this intermediate case they may still be appreciable. If so, it means that in models that reproduce the observed ice composition, stochastic effects play an important role, and further modeling of ices should rely on models that treat them properly. Since the Monte Carlo approach utilized here is very computationally demanding, although rigorous, and the modified rate equations proposed by \cite{Garrod08} are probably not entirely correct for those aspects of surface chemistry that combine two-body diffusive reactions and reverse photochemical processes (see Section 2.3), the need for further developments in mathematical methods of astrochemical modeling is clear.

\subsection{Warm-up phase}
The MONACO code allows us for the first time to examine the impact of two features of ice kinetics on the chemistry in the warm-up phase leading to hot cores: the distinction between reactive surface vs. inert bulk and the entrapment of volatiles within water ice.  We are particularly interested in how these feature affect the complex organic species that are formed in this warm-up stage. The abundances of these species in our models are shown in Table~\ref{hotcore_abu_table} along with observational values.  The model abundances refer to the time of maximum methyl formate abundance in the gas, which corresponds to 100 K. The calculated abundances of HCOOH, CH$_{3}$OCH$_{3}$ and CH$_{3}$OH are quite similar  in the Monte Carlo and RE models with $E_{\rm b}/E_{\rm D}$=0.5. The degree of agreement with observations is similar to that in \cite{GarrodHerbst06}. An order of magnitude agreement is achieved for CH$_{3}$OCH$_{3}$ in all models with $E_{\rm b}/E_{\rm D}$=0.5 and MC models with $E_{\rm b}/E_{\rm D}$=0.3 for all sources. Agreement for HCOOH is reasonable for the hot corinos IRAS~16293-2422 and IRAS~4A. For the Orion Compact Ridge and Orion Hot Core,  HCOOH is somewhat overproduced, as also occurs in \cite{GarrodHerbst06}. The abundance of gas-phase methanol after evaporation is higher in all models with $E_{\rm b}/E_{\rm D}$=0.5 and MC models with $E_{\rm b}/E_{\rm D}$=0.3 by 2--3 orders of magnitude than observed values. Overall, the results obtained with multilayer models with $E_{\rm b}/E_{\rm D}$=0.5 and 0.3 are quite similar to the results obtained with our reference model RE with $E_{\rm b}/E_{\rm D}$=0.5, and thus similar to results of \cite{GarrodHerbst06}.

Another picture is observed for methyl formate. While \cite{GarrodHerbst06} succeeded in reproducing the observed abundance of HCOOCH$_{3}$ with their two-phase RE model, our multiphase Monte Carlo models underproduce methyl formate significantly. The extent of the underproduction correlates with the number of mantle monolayers available for  surface photodissociation.  As can be seen in Figure~\ref{f12}, the abundance of HCOOCH$_{3}$ in models MC-4, where four upper monolayers of the mantle are available for surface chemistry including photodissociation, is four times higher than in models MC-1. In two-phase RE models, the entire mantle population of water is available for photodissociation, which may indicate that photoprocessing of ices is not limited to the outermost monolayers and takes place at least in a significant fraction of the bulk ice. \citet[][]{AnderssonvanDishoeck08} found from theoretical calculations that photons can penetrate about a hundred  monolayers and photodissociate species deeply embedded in the ice. Since the products of photodissociation within the bulk must somehow participate in chemical reactions, there is a  need for diffusion within the bulk at least for some species. Bulk diffusion is not included in our current Monte Carlo model and is a subject for future research.  Several currently existing models include a treatment of bulk diffusion \citep[e.g.,][]{Cuppen_ea09,Fayolle_ea11}, but to the best of our knowledge none of them has treated fully time-dependent gas-grain chemistry.

\subsection{Jumps of H$_{2}$CO abundance in the warm-up phase}
As noted above and as can be seen in Figure~\ref{f12}, the abundance profile of gas-phase formaldehyde during warm-up is more complex in multiphase Monte Carlo models than in two-phase RE models due to entrapment effects. It is interesting to see if there is any observational evidence for such a complex behavior of H$_{2}$CO. \cite{Ceccarelli_ea01} reported two jumps in the gas-phase H$_{2}$CO abundance observed around the low-mass protostar IRAS16293-2422. This object is known as a ``hot corino'' rather than a massive ``hot core'' \citep[][]{Bottinelli_ea04}. However, its temperature and density structure are similar to those adopted in our model for the warm-up stage \citep[see][]{Ceccarelli_ea00}.   In IRAS 16293-2422, the density rises from $\sim$10$^{6}$~cm$^{-3}$ at 1000~AU at the edge to $\sim$10$^{8}$~cm$^{-3}$ near the center of the object, while the temperature increases from $\sim$40~K to 150~K over the same spatial range.  According to \cite{Ceccarelli_ea01}, the first jump in abundance, from $4\times 10^{-10}$ to $4 \times 10^{-9}$, occurs at the point where the  dust temperature exceeds $\sim$50~K, the sublimation temperature of formaldehyde, while the second jump, to 10$^{-7}$, occurs at the point where the dust temperature reaches 100~K, the sublimation temperature of water ice.  According to \cite{Ceccarelli_ea00}, the gas density at the first point is $\sim 5 \times 10^{6}$~cm$^{-3}$, while at the second point it is $\sim5 \times 10^{7}$~cm$^{-3}$. For convenience, we summarize observed and modeled abundances of formaldehyde in Table~\ref{hotcore_h2co_table}.  From our models, we picked abundances of H$_{2}$CO at three times/temperatures, which from our point of view  correspond to the three shells of IRAS16293-2422 \citep{Ceccarelli_ea00}. Abundances in the column H$_{2}$CO$_{\rm hot}$ are taken from the warm-up phase of our model at time $2 \times10^{5}$~yr and correspond to the outer edge of   IRAS16293-2422, at a temperature of 200~K. Abundances in the column H$_{2}$CO$_{\rm warm}$ are taken from the warm-up phase at $1 \times 10^{5}$~yr where the temperature reaches 50~K. This point corresponds to the outer edge of the ``warm envelope'' of the source. Finally, abundances in the column H$_{2}$CO$_{\rm cold}$ are taken from the cold collapse phase of the model at the moment when the gas density reaches $5 \times 10^{6}$~cm$^{-3}$. These abundances correspond to the ``cold envelope'' of IRAS16293-2422 \citep{Ceccarelli_ea01}.

One can see that the multiphase models MC-1 and MC-4 with $E_{\rm b}/E_{\rm D}$=0.3 and $E_{\rm b}/E_{\rm D}$=0.5 qualitatively reproduce the observed H$_{2}$CO abundance distribution for IRAS 16293-2422 \citep{Ceccarelli_ea01}. Specifically, the modeled abundances of formaldehyde at cold and warm regions agree with \cite{Ceccarelli_ea01} within an order of magnitude. On the other hand, the modeled H$_{2}$CO abundance  at the outer edge of the hot corino at 200~K is  too high by 1--2 orders of magnitude. As compared with the multiphase models,  the two-phase  RE models do not reproduce the observations of \cite{Ceccarelli_ea01} as well, especially the model with $E_{\rm b}$/$E_{\rm D} = 0.5$. The abundance of formaldehyde in the warm envelope is determined by the evaporation of H$_{2}$CO from the ice surface while the majority of the species remains locked within water ice. Two-phase models cannot account for this effect. The exception is our RE model with $E_{\rm b}$/$E_{\rm D}$=0.3. However, this model cannot be considered as successful, because it produces an anomalous ice composition, as seen in Figure~\ref{rebestfit}. Note that when analyzing the warm H$_{2}$CO abundance, we ignore for two reasons the sharp spike that can be seen at 8$\times$10$^{4}$ yr in Figure~\ref{f12} in the MC-1 and MC-4 models.  First, this spike occurs at a somewhat lower temperature than claimed in \cite{Ceccarelli_ea01} for the warm envelope.  Second, the lifetime of this spike is very small and it should not be seen in observations.

In the later study of \cite{Maret_ea04}, the H$_{2}$CO emission was studied towards eight other sources. In four of them (NGC1333-IRAS~4B, NGC1333-IRAS~2, L1448-MM and L1527), the fitting of observational data is  best when the existence of a jump in the formaldehyde abundance at a temperature near 100~K is assumed. However, a second jump near 50~K, as found in IRAS~16293-2422 by \cite{Ceccarelli_ea01}, was not confirmed for these sources due to limitations in observational data. Abundances of H$_{2}$CO derived from the fitting of observational data in \cite{Maret_ea04} for these four sources are listed in the last four rows of Table~\ref{hotcore_h2co_table}.  Again, the observed abundances of formaldehyde are  more consistent with the results of multiphase than two-phase modeling.  Specifically, only in multiphase models is the majority of H$_{2}$CO  released to the gas from ice mantles  at temperatures near 100~K, as seen in Figure~\ref{f12}.

\section{Summary}\label{summary}
In this study, we constructed a multiphase gas-grain Monte Carlo model of chemical evolution in the interstellar medium. The model treats icy grain mantles in a layer-by-layer manner. The ice treatment includes  distinctions between the reactive outer monolayers of an ice mantle and the inert bulk, as well as between zeroth and first order desorption.  It also includes entrapment of volatile species in a refractory ice mantle. One advantage of our multiphase model over previously considered three-phase models is that it is capable not just of accreting, but also of desorbing the ice mantle in a layer-by-layer manner, which is important for warm-up chemistry. The behavior of ice during desorption in our model is justified by comparison with laboratory TPD data on ice desorption.

With the  MONACO model, the chemical evolution during the cold and warm-up phases leading to hot core/corino formation has been investigated. For the first time, the impact of a detailed multilayer approach to grain-mantle formation on the warm-up chemistry has been explored. Overall, our model for the first time simultaneously explains the observed composition of ice mantles formed during the cold stage of prestellar evolution, and abundances of gas-phase species evaporated from the ice during the warm-up protostellar stage.

We find that the cooling down of dust from 20~K to 10~K during the cold collapse phase, which is caused by the shielding of interstellar radiation during contraction from the diffuse to dark cloud phase, helps us to explain the observed ice composition and structure semi-quantitatively. Multiphase Monte Carlo models reproduce the observed ice composition better than two-phase models. Also, contrary to two-phase models, multiphase models explain the observed ice structure, including a polar mixture of CO$_{2}$ and water at monolayers close to the bottom of the mantle, and an apolar phase of ice rich in CO close to the top of the mantle.

The impact of the multilayer ice-modeling approach on the abundances of organic species formed during the warm-up phase is somewhat mixed. Species such as HCOOH and CH$_{3}$OCH$_{3}$ are  affected only slightly by the new approach, because their abundances to a large extent are controlled by gas-phase reactions. On the contrary, the abundance of HCOOCH$_{3}$ is different in two-phase and multiphase models. Here, the treatment of photochemistry is crucial for the modeled abundance of HCOOCH$_{3}$. Unlike two-phase models, in which photodissociation occurs throughout the mantle, our multiphase Monte Carlo approach allows photodissociation in only the top four monolayers of the mantle.  This restriction leads to an underproduction of gaseous methyl formate compared with observation. The failure of the MC models might be an indication that  photochemistry and bulk diffusion could be active deep in the bulk ice and that some diffusion of the products to the ice surface exists in real ices.

Contrary to two-phase models, our multiphase modeling explains jumps at 50~K and 100~K in the H$_{2}$CO abundance found by \cite{Ceccarelli_ea01} in the hot corino IRAS 16293-2422, and jumps of H$_{2}$CO abundance at 100~K observed in other sources by \cite{Maret_ea04}. The 100~K jumps are explained as due to release of formaldehyde from the refractory water ice mantle at the sublimation temperature of H$_{2}$O. This effect cannot be explained with two-phase models because it requires a consideration of the partial entrapment of formaldehyde in refractory water ice at 50~K. An analogous two-step behavior of gas-phase species in our multiphase model is predicted to occur for other volatile species such as CO, CH$_{4}$, and CO$_{2}$, and may be a subject for further observational studies of star formation regions.

Various parameters controlling the grain surface chemistry have a different impact on the resulting ice composition. The diffusion/desorption energy ratio $E_{\rm b}/E_{\rm D}$ significantly affects the modeled ice composition. The best agreement between the modeled ice composition and observations is achieved in models with $E_{\rm b}/E_{\rm D}$=0.5.  Models with $E_{\rm b}/E_{\rm D}$=0.3 are also reasonable for the warm-up phase, while the ice composition produced with these models is somewhat different from observations. We included neither the competition between diffusion and desorption competition, nor the mechanism of oxygen hydrogenation on top of a buried CO molecule, as described in \cite{GarrodPauly11}.  Nevertheless, our results are similar to the results of that study for the cold ice phase.  One reason for this agreement, as regards CO$_{2}$ ice, is that the reaction ${\rm grCO}+{\rm grOH}\rightarrow {\rm grCO_{2}}+{\rm H}$ at the slightly higher temperatures we use can form CO$_{2}$ on cold surfaces in sufficient amounts.

We have used two different sets of stochastic models - MC-4 and MC-1.  Models MC-1 are obtained by applying corrections (\ref{rscorr}) and (\ref{rscorrph}) to model MC-4 (see Section 2.3.1).     Correction (\ref{rscorr}) is related to the possibility of reactions between species residing in different monolayers of ice, while correction (\ref{rscorrph}) is related to the deepness of penetration of photons into the ice mantle.
The use of models MC-1 leads to better agreement with the observed ice composition at the cold collapse phase than achieved with models  MC-4. However, at higher temperatures, as occur during the warm-up stage, the two sets of models show little difference except that MC-4 exhibits better agreement with the observed abundance of methyl formate.

Overall, this work shows that realistic multiphase macroscopic Monte Carlo models of gas-grain interstellar chemistry are not only feasible with modern computers, but also better explain observational data than models based on two phases and rate equations.

\acknowledgements
AV wishes to thank Dr. Qiang Chang for useful discussions on the mobility of species on rough surfaces.  EH wishes to  acknowledge the support
of the National Science Foundation for his astrochemistry program. He also acknowledges support from the NASA Exobiology and Evolutionary Biology program through a subcontract from Rensselaer Polytechnic Institute. The authors would like to thank the anonymous referee for her/his valuable comments. This research has made use of NASA's Astrophysics Data System.


\clearpage

\appendix

\section{The MONACO algorithm}
In this section, a detailed description of the MONACO algorithm is given. Since the treatment of gas-phase chemistry in the algorithm exactly follows that described in \cite{Gillespie76}, below we focus on the steps pertaining to surface chemistry. Nevertheless, a step-by-step description of the algorithm given below includes all steps, both for gas-phase and grain surface chemistry.

\begin{figure}\centering
\includegraphics[width=0.8\textwidth]{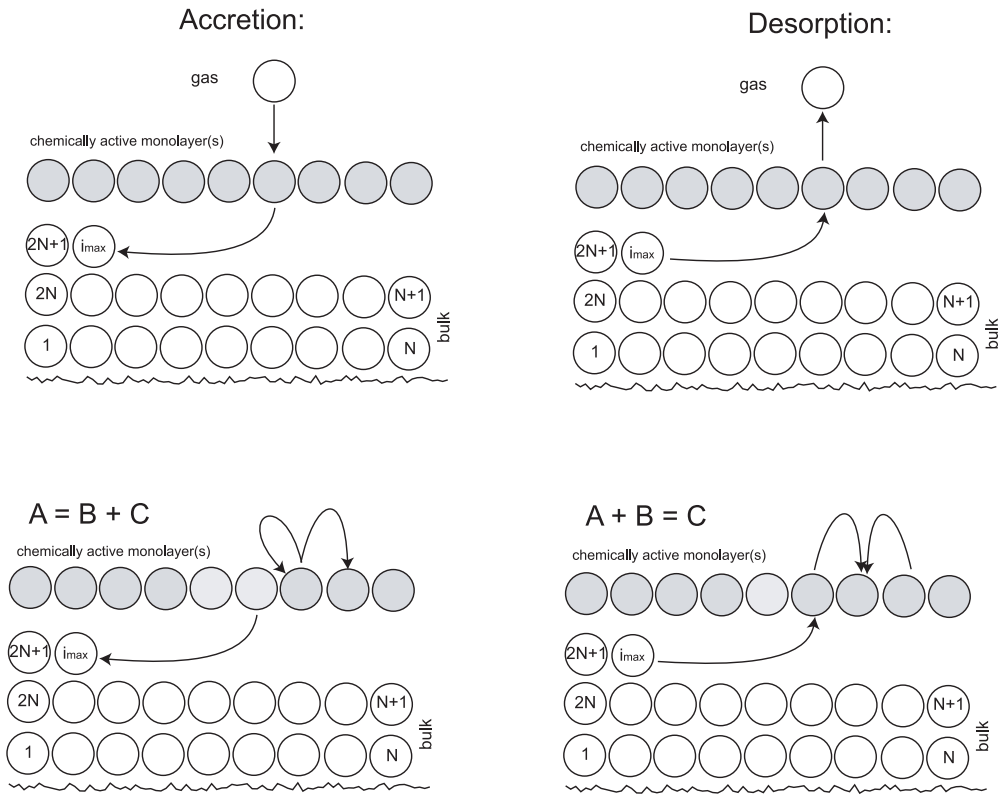}\caption{Processes occurring on an ice mantle in which the outermost layer is full in the MONACO model.  Excess species adsorbed into this layer or produced on it by reaction lead to the random addition of an excess species to a virtual queue, shown beneath the outermost layer.  The virtual queue is reduced in its number of species when desorption occurs from the outermost layer.}\label{monaco_scheme}
\end{figure}

In Figure~\ref{monaco_scheme}, the representation of a multilayer ice mantle in the MONACO algorithm is shown along with assorted processes. Since a dust particle has a finite size and finite number of adsorption sites $N_{\rm s}$, only a finite number $N_{\rm s}$ of molecules can be placed on it in one monolayer. If the total number of molecules $M_{\rm s}$ on a grain does not exceed $N_{\rm s}$, the first, or outermost, monolayer is forming.  The first monolayer is considered to be chemically active; i.e., accretion, desorption, and chemical reactions on a grain can proceed only there.  If $M_{\rm s}$ exceeds $N_{\rm s}$, a molecule is randomly picked from the first layer and moved to an ordered queue with index $i$ starting from 1. The next removed molecule will have an index $i+1$ and so on. If, due to desorption or a surface reaction with fewer products than reactants, the number of molecules on the outermost layer is reduced, the molecule with the highest index $i_{\rm max}$ is  moved back onto the outermost layer and $i$ is lessened by 1. Therefore, when $M_{\rm s}\ge N_{\rm s}$, the total surface population of molecules in the outermost monolayer is always kept at $N_{\rm s}$ while the number of molecules in the queue changes with time according to the chemical evolution of the system.  In contrast to the chemically active monolayer, molecules in the queue are chemically inert;  i.e., do not participate in chemical reactions, accretion or desorption processes.

We consider the spans of $i_{\rm max}~...~i_{\rm max}-N_{\rm s}$, $i_{\rm max}-N_{\rm s}~...~i_{\rm max}-2N_{\rm s}$ etc. as the Mth bulk layer, (M-1)th bulk layer and so on. The layer with number 1 consists of molecules that came to the bulk at the beginning of system time. The last bulk layer, with the highest number, contains the most recent molecules buried in the bulk. Therefore, one can consider these model layers that together form the queue as a representation of layers in a physical mantle below the outermost layer. A step-by-step description of the algorithm is given below. The following terms are used. $a_i$ is a rate of $i$-th chemical reaction. $\vec{S}=(S_1,S_2,...,S_{\rm N},S_1^s,S_2^s,...,S_{\rm N}^s)$ is a vector of abundances of all species; $S_{\rm j}$ is an abundance of $j$-th species in gas phase, $S_{\rm j}^s$ in an abundance of $j$-th species on a grain,  $\vec{\nu_k}$ is the stoichiometry vector for reaction $k$, which shows how the abundances of reactants and products of reaction $k$ change due to the single event of the reaction,  $t_{\rm current}$ is the current physical time of a simulated system, $t_{\rm end}$ is the time until the chemical evolution of the system is finished, and $N_{\rm tot}^{max}$ is the maximum number of molecules in the chemically active monolayer. $N_{\rm tot}$ is the actual number of molecules in chemically active monolayers.  By setting $N_{\rm tot}^{max}$ to be larger than $N_{\rm s}$, we assume that species in more than one physical monolayer of ice are chemically active. In this study, we use $N_{\rm tot}^{max}=4\times N_{\rm s}$ based on the results of Section~\ref{labbeh}.  Finally, Q and q$_{max}$ are defined as arrays representing the ordered queue and the number of molecules in the queue, correspondingly. All processes in the model are treated as unimolecular or bimolecular reactions. In particular, accretion and desorption are represented as unimolecular chemical reactions: ${\rm X}\rightarrow {\rm grX}$, ${\rm grX}\rightarrow {\rm X}$.

The algorithm steps are as follows:

\begin{itemize}
\item
Set initial abundances and the queue: $\vec{S}=\vec{S_0}$, Q = Q$_0$, $q_{\rm max}=q_{\rm max}^0$.
\item
While $t_{\rm current} \le t_{\rm end}$:
\begin{enumerate}
\item\label{BeginGillespie}
Calculate rates of all reactions $a_i$
\item
Calculate the sum of rates of all reactions: $a_0 = \sum_{i=1}^{i=N} a_i$
\item
Choose uniformly distributed random numbers r$_{1}$ and $r_{2}$
\item
Calculate time step: $\Delta t = \frac{1}{a_0}\cdot ln(\frac{1}{r_1})$, update $t_{\rm current} = t_{\rm current} + \Delta t$
\item\label{EndGillespie}
Choose process $k$ to happen: $k = min: \sum_{i=1}^{i=k}a_i < a_0\cdot r_2$ and update abundances: $\vec{S} = \vec{S} + \vec{\nu_k}$
\item\label{BeginOurs}
If $k$ alters surface species, update total number of molecules in the chemically active monolayers
$N_{\rm tot}$
\item
If $N_{\rm tot} > N_{\rm tot}^{max}$:
\begin{enumerate}
\item
Choose uniformly distributed random number $r_{3}$
\item
Choose molecule $m$ to be buried in the bulk: $m = min: \sum_{j=1}^{j=m}S_i^s <
N_{\rm tot}\cdot r_3$ and update abundances: $S_{\rm j}^s = S_{\rm j}^s - 1$, $q_{\rm max} = q_{\rm max} + 1$, $Q(q_{\rm max}) = m$
\end{enumerate}
\item\label{EndOurs}
If $N_{\rm tot} < N_{\rm tot}^{max}$:
return molecule $r$ from bulk to surface: $r = Q(q_{\rm max})$, $S_r^s = S_r^s + 1$, $q_{\rm max} = q_{\rm max} - 1$
\end{enumerate}
\item
If $t_{\rm end}$ is reached, stop the simulation and save results, else go to step \ref{BeginGillespie}.
\end{itemize}

Steps \ref{BeginGillespie}--\ref{EndGillespie} represent the classical Gillespie algorithm exactly \citep{Gillespie76}, while steps \ref{BeginOurs}--\ref{EndOurs} are the addendum developed in this study to account for the ice composition in the framework of the macroscopic Monte Carlo approach.

\clearpage

\begin{table*}
\caption{Best fit ice composition versus observed ice composition towards low-mass protostars$^{a}$.}\label{ice_comp_table}
\begin{tabular}{l|c|c|cccccc}
\hline\hline
Model & Best fit time (yr) & Min $F_{\rm fit}$& H$_{2}$O & CO & CO$_{2}$ & CH$_{3}$OH & NH$_{3}$ & CH$_{4}$ \\
\hline
RE, $E_{\rm b}/E_{\rm D}$=0.3    & 8.0$\times$10$^{5}$ & 1.10 & 100 & 4  & 170 & 2  & 14 & 8  \\
RE, $E_{\rm b}/E_{\rm D}$=0.5    & 4.0$\times$10$^{5}$ & 0.02 & 100 & 41 & 41  & 0  & 14 & 9  \\
RE, $E_{\rm b}/E_{\rm D}$=0.77   & 4.7$\times$10$^{5}$ & 0.11 & 100 & 49 & 0   & 0  & 14 & 9  \\
MC-1, $E_{\rm b}/E_{\rm D}$=0.3  & 9.7$\times$10$^{5}$ & 0.20 & 100 & 11 & 22  & 34 & 25 & 14 \\
MC-1, $E_{\rm b}/E_{\rm D}$=0.5  & 8.3$\times$10$^{5}$ & 0.03 & 100 & 42 & 24  & 5  & 21 & 9  \\
MC-1, $E_{\rm b}/E_{\rm D}$=0.77 & 6.3$\times$10$^{5}$ & 0.19 & 100 & 66 & 2   & 0  & 21 & 8  \\
MC-4, $E_{\rm b}/E_{\rm D}$=0.3  & 9.8$\times$10$^{5}$ & 0.21 & 100 & 10 & 33  & 35 & 24 & 13 \\
MC-4, $E_{\rm b}/E_{\rm D}$=0.5  & 9.8$\times$10$^{5}$ & 0.07 & 100 & 30 & 19  & 22 & 20 & 10 \\
MC-4, $E_{\rm b}/E_{\rm D}$=0.77 & 4.5$\times$10$^{5}$ & 0.15 & 100 & 54 & 4   & 0  & 16 & 8  \\\hline
Observational value        & ---                 & ---  & 100 & 29 & 29  & 3  & 5  & 5  \\
\hline
\end{tabular}
\tablenotetext{a}{ Observational data are taken from Table~2 of \cite{Oeberg_ea11}.}
\end{table*}

\begin{table}
\caption{Modeled peak abundances of selected gas-phase species in the cold collapse phase versus observational values for TMC-1(CP)$^{a}$.}\label{gas_phase_species}
\begin{tabular}{lcc}
\hline\hline
Species & Model & Observations \\
\hline
H$_{2}$O & 2(-7) & $\le$7(-8) \\
HCO$^{+}$ & 4(-9) & 8(-9) \\
NH$_{3}$ & 3(-8) & 2(-8) \\
N$_{2}$H$^{+}$ & 1(-9) & 4(-10) \\
H$_{2}$CO & 2(-8) & 5(-8) \\
H$_{2}$S & 3-9(-11) & 5(-10) \\
\hline
\end{tabular}
\tablenotetext{a}{Observational abundances are taken from the compilation in \cite{Smith_ea04}.}
\end{table}

\clearpage

\begin{deluxetable}{l|rrrrr}
\tablecolumns{6}
\tablewidth{0pc}
\tablecaption{Modeled and observed abundances of gaseous complex species in hot cores/corinos with respect to $n_{\rm H}${$^{a}$}.}
\tablehead{
\colhead{Model} & \colhead{HCOOCH$_{3}$-A} & \colhead{HCOOCH$_{3}$-E} & \colhead{HCOOH} & \colhead{CH$_{3}$OCH$_{3}$} & \colhead{CH$_{3}$OH}
}
\startdata
RE, $E_{\rm b}/E_{\rm D}$=0.3    & 2.7(-8)  & & 2.1(-9)  & 7.3(-10) & 5.3(-7) \\
RE, $E_{\rm b}/E_{\rm D}$=0.5    & 2.0(-8)  & & 5.2(-8)  & 4.3(-8)  & 2.0(-5) \\
RE, $E_{\rm b}/E_{\rm D}$=0.77   & 7.9(-11) & & 5.8(-10) & 3.6(-12) & 4.7(-8) \\
MC-1, $E_{\rm b}/E_{\rm D}$=0.3  & 3.9(-10) & & 4.2(-8)  & 6.3(-8)  & 4.1(-5) \\
MC-1, $E_{\rm b}/E_{\rm D}$=0.5  & 8.0(-11) & & 1.2(-8)  & 1.2(-8)  & 4.8(-6) \\
MC-4, $E_{\rm b}/E_{\rm D}$=0.3  & 1.2(-9)  & & 8.9(-8)  & 6.4(-8)  & 3.7(-5) \\
MC-4, $E_{\rm b}/E_{\rm D}$=0.5  & 3.4(-10) & & 1.4(-8)  & 4.3(-8)  & 2.5(-5) \\
\hline
IRAS 16293-2422                 & 8.5$\pm$0.4(-8) & 1.2$\pm$0.4(-7) & 3.1(-8)  & 1.2(-7)      & 5.0(-8)      \\
IRAS 4A                    & 1.7$\pm$0.9(-8) & 1.8$\pm$0.9(-8) & 2.3(-9)  & $\le$1.4(-8) & $\le$5.0(-9) \\
Orion Compact Ridge        & 1.5(-8)         & ---             & 7.0(-10) & 9.5(-9)      & 2.0(-7)      \\
Orion Hot Core             & 7.0(-9)         & ---             & 4.0(-10) & 4.0(-9)      & 7.0(-8)      \\
\enddata
\tablenotetext{a}{Models MC-1 and MC-4 with $E_{\rm b}/E_{\rm D}$=0.77 are excluded because they produce very little HCOOCH$_{3}$. Modeled abundances are chosen at 100 K, which is the temperature at which gaseous methyl formate is at its maximum abundance.  The abundances of all species in IRAS 16293-2422 and IRAS 4A except for HCOOCH$_{3}$-E are taken from \cite{Bottinelli_ea07}, while the abundances of HCOOCH$_{3}$-E in these sources are taken from \cite{Bottinelli_ea04}. The abundances for the Orion Compact Ridge and Hot Core are taken from \cite{Sutton_ea95}.}
\label{hotcore_abu_table}
\end{deluxetable}

\clearpage

\begin{deluxetable}{l|rrr}
\tablecolumns{4}
\tablewidth{0pc}
\tablecaption{Modeled and observed abundances of gaseous H$_{2}$CO in hot cores/corinos with respect to $n_{\rm H}${$^{a}$}.}
\tablehead{
\colhead{Model} & \colhead{H$_{2}$CO$_{\rm hot}$} & \colhead{H$_{2}$CO$_{\rm warm}$} & \colhead{H$_{2}$CO$_{\rm cold}$} }

\startdata
RE, $E_{\rm b}/E_{\rm D}$=0.3    & 4.2(-8)  & 7.8(-10) & 5.2(-11) \\
RE, $E_{\rm b}/E_{\rm D}$=0.5    & 6.2(-7)  & 1.6(-5)  & 1.7(-11) \\
RE, $E_{\rm b}/E_{\rm D}$=0.77   & 2.6(-12) & 1.3(-8)  & 3.1(-11) \\
MC-1, $E_{\rm b}/E_{\rm D}$=0.3  & 1.7(-6)  & 3.1(-9)  & 3.2(-11) \\
MC-1, $E_{\rm b}/E_{\rm D}$=0.5  & 9.7(-7)  & 1.2(-9)  & 5.0(-11) \\
MC-1, $E_{\rm b}/E_{\rm D}$=0.77 & 2.7(-12) & 3.7(-10) & 3.0(-11) \\
MC-4, $E_{\rm b}/E_{\rm D}$=0.3  & 1.6(-6)  & 1.0(-8)  & 4.0(-11) \\
MC-4, $E_{\rm b}/E_{\rm D}$=0.5  & 2.7(-6)  & 4.1(-9)  & 3.4(-11) \\
MC-4, $E_{\rm b}/E_{\rm D}$=0.77 & 4.8(-12) & 4.6(-10) & 3.7(-11) \\
\hline
IRAS 16293-2422                 & 5.0(-8) & 2.0(-9)  & 2.0(-10) \\
IRAS 4B                    & 1.5(-6) & 2.5(-10) & ---      \\
IRAS 2                     & 1.0(-7) & 1.5(-10) & ---      \\
L1448-MM                   & 3.0(-7) & 4.0(-11) & ---      \\
L1527                      & 3.0(-6) & 1.5(-10) & ---      \\
\enddata
\tablenotetext{a}{Abundance of H$_{2}$CO in IRAS 16293-2422 is taken from \cite{Ceccarelli_ea01}. Abundances for other sources are taken from \cite{Maret_ea04}.}
\label{hotcore_h2co_table}
\end{deluxetable}

\clearpage

\clearpage

\clearpage

\clearpage

\clearpage

\end{document}